\documentclass[aps,preprint]{revtex4}
\usepackage{amsmath}
\usepackage{amssymb}
\usepackage{graphicx}

\begin{document}

\title{Maximum entropy states of collisionless systems with long-range
interaction and different degrees of mixing}
\author{V. M. Pergamenshchik}
\email{victorpergam@yahoo.com}
\affiliation{Institute of Physics, prospect Nauki 46, Kiev 03039, Ukraine}
\date{\today }

\begin{abstract}
Dynamics of many-particle systems with long-range interaction is
collisionless and governed by the Vlasov equation. This dynamics is a flow
of a six-dimensional incompressible liquid with uncountable integrals of
motion. If the flow possesses the statistical property of mixing, each
liquid element spreads over the entire accessible space. I derive the
equilibrium microcanonical maximum entropy states of this liquid for
different degrees of mixing $M$. This $M$ is the number of liquid elements
which are statistically independent. To count microstates of a liquid, I
develop analog of the discrete combinatorics for continuous systems by
introducing the ensemble of phase subspaces and making contact with the
Shannon-McMillan-Breiman theorem from the ergodic theory. If $M$ is much
larger than the total number of particles $N$, then the equilibrium
distribution function (DF) is found to be exactly of the Fermi-Dirac form.
If the system is ergodic but without mixing, $M=0$, the DF is a formal
expression which coincides with the famous DF obtained by Lynden-Bell. If
the mixing is incomplete and $M\sim N$, the exponentials, which are present
in Lynden-Bell's DF, appear with certain weight given by the entropy of
mixing. For certainty, the \ long-range interaction is taken in the form of
the Newton and Coulomb potential in three dimensional space, but the
applicability of the method developed in the paper is not restricted to this
case. The analogy of the obtained statistics to the Fermi-Dirac statistics
allows for expressing the entropy of the system via its total energy and
chemical potentials of liquid's elements. The effect of the long-range
interaction to the basic thermodynamic relations is demonstrated.\newline

PACS: 5.20.-y - Classical statistical mechanics; 05.70.-a - Thermodynamics.

Keywords: Long range interaction, Vlasov's 6D liquid, mixing, entropy
\end{abstract}

\maketitle

\section{Introduction}

\subsection{Lynden-Bell's maximum entropy state of collisionless systems}

Interaction potentials which decay not more rapidly than $1/R^{D},$ where $D$
is the space dimensionality and $R$ is the particles' separation, are
considered as long range ones. The reason is that for a long-range
potential, the interaction of a particle with the more remote particles is
stronger than that with less remote particles, so that the interaction with
closest neighbors in sufficiently large systems becomes negligible. As to a
good approximation remote particles constitute a continuous medium, the
interaction is determined by their spatially averaged density which is the
source of a mean field. Hence systems with long-range interaction can be
well-represented by their smooth densities and mean fields disregarding the
effects of matter discreetness and, in particular, pair encounters. This is
exactly the case of applicability of the collisionless Vlasov equation which
describes evolution of a smooth distribution function (DF) in the
six-dimensional (6D) space of particles' coordinates and velocities under
the action of the self-consistent mean field. This 6D space is called $\mu $
space. Thus, dynamics of many-particle systems with long-range interaction
in the $\mu $ space is collisionless and governed by the Vlasov equation
\cite{Phys Rep 1990,book,Phys Rep 2009,Phys Rep 2014}.

The final state to which any large classical system relaxes is the standard
Maxwell-Boltzmann equilibrium as pair collisions will eventually manifest
themselves. But in systems with long-range interaction the time when this
happens increases approximately as the number of particle to the power which
is larger than unity \cite{1.7 1,1.7 2,1.7 3,1.7 4,1.7 5,1.7 6} so that for
a very long time the system is governed by the collisionless Vlasov
equation. As a result, the physics of many long-range systems is essentially
collisionless. For instance, the relaxation time due to pair collisions in
galaxies is estimated to be much longer than the age of our Universe (which
is known as Zwicky paradox). At the same time, the remarkable regularity in
the characteristics of \ the gravitational systems, such as galaxies,
suggests that these systems have reached some equilibrium states. This was
the principal motivation for Lynden-Bell to consider the statistical
equilibrium state to which a collisionless gravitational system can relax in
its own violently changing mean field \cite{LB}. The main challenge of this
collisionless statistical approach was that the Vlasov equation describes
not particles, but a 6D incompressible liquid, which is a continuous system
with the infinite number of conserved quantities. Lynden-Bell modeled the
liquid by dividing it into small stable macroparticles and then incorporated
the incompressibility by not allowing two macroparticles to occupy the same
cell of the $\mu $ space. Using the standard combinatorics for the discrete
macroparticles, Lynden-Bell arrived at his famous Fermi-like DF. This
Lynden-Bell's DF (LBDF) successfully explained the two fundamental
experimental facts which cannot be explained by the standard
Maxwell-Boltzmann statistics. First, thanks to the smooth macroscopic
character of the Vlasov equation which is insensitive to discrete properties
of individual stars, the LBDF does not discriminate between stars with
different masses as it depends on the average mass density. Second, thanks
to the incompressibility of Vlasov's liquid, the LBDF possesses the
Fermi-like degeneracy which naturally explains the dense core-rarefied galo
structure of galaxies and which, as eventually turned out, is universal for
all systems with long-range interaction \cite{Phys Rep 2014}. The impressive
elegancy of Lynden-Bell's result, the revealed evidence that the Vlasov
equation might have so a unique role in the statistical physics and its
relation to the generic systems with long-range interaction had resulted in
a wide interest to the equilibrium and dynamical properties of collisionless
systems. These studies have developed into the novel branch of statistical
physics identified as the statistical mechanics of systems with long-range
interaction \cite{Phys Rep 1990,book,Phys Rep 2009,Phys Rep 2014}.

By now it has been well established that the assumption that the violent
relaxation results in certain equilibrium maximum entropy state is an
idealization. The collisionless relaxation is incomplete because the
efficiency of the mean-field-driven mixing is dying out as the system is
approaching states with a smoother and smoother field, so that the state
with complete mixing is not attained. In reality, the relaxation can cease
at some of dynamically stable states of the Vlasov equation which are
numerous and substantially differ from the maximum entropy state \cite{1.7
4, Fignja PRL,Phys Rep 2014}. Nevertheless, since its prediction by
Lynden-Bell in 1967 for stellar systems, the equilibrium statistical
mechanics of this idealized state has been attracting an unrelenting
attention of theoretical physicists. The result obtained by Lynden-Bell has
been rederived by various technics \cite{Phys Lett,Tech 1,Tech 2,Tech 3,Tech
3a,Tech 4,Chav book}, praised and criticized \cite{LB critics}, its
quantitative predictability has been supported in some cases \cite{Phys Rep
2014,Prais 1,Prais 2,Prais 3} and denied in other cases \cite{Critics
2,Critics 3}, its validity has been reconsidered and a different form of the
equilibrium state has been proposed \cite{Nakamura}. One reason is that
systems with long-range interaction, hence collisionless and related with
the LB statistics, appear in many different fields of physics: gravitational
systems in astrophysics, charged beams with Coulomb interaction in the
plasma physics, two-dimensional hydrodynamics where the interaction of
vortices is logarithmic, systems of many spherical colloids at liquid
surfaces where the attraction scales as $1/R^{2}$ \cite{PRE2009}, dipolar
systems and ferroelectric liquids. Another reason is that the thermodynamics
of systems with long-range interaction cannot be based on the fundamental
axioms of the standard thermodynamics of systems with short-range
interactions such as spatial uniformity, additivity of entropy and energy,
equivalence of the canonical and microcanonical description, presence of a
thermal bath fixing the system's temperature \cite{Phys Rep 1990,book,Phys
Rep 2009,Phys Rep 2014}. This brings a fresh breath to the thermodynamics.
There is also a very peculiar source of attraction, a challenge unformulated
explicitly but making theorists think of collisionless maximum entropy
states which under some, even highly idealized assumptions, can result from
the collisionless Vlasov dynamics.

The problem is that, as mentioned above, the Vlasov equation describes
motion of a 6D incompressible liquid. But, on the one hand, neither a liquid
can be approximated by a set of separated shape-preserving fragments, nor
any liquid fragment preserves its shape. This is true for any motion, but is
particularly fundamental for a chaotic mixing motion which can possibly
result in a maximum entropy state in any sense: such \ a motion is supposed
to eliminate correlations between initially neighboring liquid elements
which implies producing thinner and thinner liquid filaments. On the other
hand, while a verity of technics has been employed to address collisionless
statistical equilibrium in different systems with long-range interaction,
the common device for counting the number of states of the Vlasov liquid has
been either similar or essentially equivalent to the original Lynden-Bell's
device, i.e., dividing the liquid into independent shape-preserving
fragments called macroparticles, dividing the $\mu $ space into cells each
of which can accommodate exactly one macroparticle or nothing, and using the
discrete combinatorics for counting possible distributions of the
macroparticles over these cells \cite{Phys Lett,Tech 1,Tech 2,Tech 3,Tech
3a,Tech 4,Chav book}. The introduction of macroparticles allows one to
substitute the problem of the innumerable liquid degrees of freedom for the
well-known problem of a finite number of particles, but the very idea of a
nonspreading macroparticle contradicts the idea of the mixing dynamics which
can lead to any kind of statistical equilibrium. Indeed, in a mixing flow
any small liquid element spreads over the whole accessible space so that its
filaments can enter any cell of the $\mu $ space. We see that the main
disadvantage of the present theory is the fundamental contradiction between
the idea of solid, undiffuse and independent macroparticles and the
continuity of a mixing flow of the 6D liquid.

The problem of finding a maximum entropy state or, in other words, a
statistical equilibrium state, which can result from the Vlasov equation
under highly idealized assumptions on the chaotic dynamics, may be described
as follows. First of all, one has to deal with an actual 6D liquid rather
than with any kind of particles (i.e., not with real particles,
macroparticles, vortices). In the statistical equilibrium all the accessible
states are equally probable. But any change of liquid's shape results in its
new state. The question is how to describe the macroscopic and microscopic
states of this continuous system and how to count its states in order to
obtain the entropy. Moreover, it is known that the Vlasov system possesses
the uncountable number of conserved quantities, values of the DF at each
point of the liquid usually called levels, which do not change along their
trajectories. These integrals of motion must be imposed as certain
restrictions on the accessible states. It is these integrals that account
for the incompressibility and the related Fermi-like exclusion property.
However, in contrast to the simple "one cell-one macroparticle" restriction
of Lynden-Bell's model, the actual incompressibility can only mean that the
total amount of liquid with all different levels in a given cell of the $\mu
$ space \textit{cannot exceed its 6D volume} $\omega $. Hence the amount of
liquid in a given cell, which is either zero or one in Lynden-Bell's model,
is a \textit{continuous number} ranging from $0$ to $\omega $, so that the
summation over the occupation numbers must be replaced by some integration.
Finally, in principle, the degree of mixing of the 6D liquid can be
different, and one has to specify this degree and find its connection to the
final equilibrium state.

In this paper I present the theory of the maximum entropy state of a 6D
liquid to which systems with a long-range interaction can relax under the
assumption of the chaotic dynamics governed by the Vlasov equation. It turns
out that the solution to this problems is not unique as it depends on the
presupposed statistical properties of the Vlasov dynamics. According to the
ergodic theory \cite{Balescu,Reichl,Bil,Sinai}, the chaoticity of a
dynamical system ranges from ergodicity without mixing, which is the lowest
statistical property, through mixing with zero Kolmogorov-Sinai entropy to
the so-called K-mixing with nonzero Kolmogorov-Sinai entropy, which is the
highest statistical property. In the next Section I present some information
from the ergodic theory which is instructive to our problem, give further
details of the approach, and briefly describe the structure and results of
this paper.

\subsection{Possible statistical properties of a flow of liquid: ergodicity,
mixing, and incomplete mixing}

Motion of a liquid in any space is a flow. Flows on different phase spaces
and their statistical properties are the subject of the ergodic theory \cite%
{Balescu,Reichl,Bil,Sinai}. It is therefore natural to resort to the ideas
of the ergodic theory for developing a statistical approach to the 6D Vlasov
liquid. One of the most important quantities in the language of the ergodic
theory is the measure (volume) $\mu (A\cap B)$ of intersection $A\cap B$ of
a set $A$ with a set $B.$ This quantity determines probability that a point
from $A$ can be found in the set $B$ and thus defines the coarse-grained DF
of $A$. The flow implies that the set $A$ occupied by liquid changes its
shape and is a function of time $t$, i.e., $A=A(t)$. The quantity $\mu
(A(t)\cap B)$ is adequate to specify the shape $A(t)$ of a flowing set $A$
at an instant $t$ with respect to the fixed sets $\{B\}$ which can be used
as a reference frame for a coarse-grained description of the flow $A(t).$ At
last, the fundamental statistical properties of dynamical systems in the
ergodic theory can be expressed in terms of the above measure of
intersection.

Ergodic and mixing flows preserve the volume of a flowing set: $\mu
(A(t))=\mu (A(0))=\mu (A).$ Ergodicity is the weakest statistical property
of a dynamical system which does not imply its relaxation to some stationary
state. In an ergodic (but non-mixing) flow $A(t)$ of a set $A$, its measure
of intersection with a fixed set $B$ remains time dependent at all times.
The only time independent quantity is the temporal average of $\mu (A(t)\cap
B)$ \cite{Sinai}:
\begin{equation}
\underset{t\rightarrow \infty }{\lim }\frac{1}{t}\int_{0}^{t}\mu (A(\tau
)\cap B)d\tau =\mu (A)\mu (B).  \label{Erg}
\end{equation}%
Hence, measuring the amount of liquid $A$ in the fixed set $B$, $\mu
(A(t)\cap B)$, at different times $t$ will give different results as $\mu
(A(t)\cap B)$ as a function of $t$ is not converging to any stationary
value. We see that the ergodicity alone does not result in a relaxation. The
formula (\ref{Erg}) tells one that in an ergodic flow, a liquid fragment $A$
does not spread over the accessible space, its points do not decorrelate
and, as time elapses, $A$ moves roughly as a single piece. Thus, the
chaoticity of the ergodic motion of $A$ consists in the fact that $A$ can be
found everywhere with equal probability.

A mixing flow is ergodic, but converse is not necessarily true. Mixing is a
much stronger statistical property than ergodicity and can be naturally
associated with a relaxation towards certain time independent state. In a
mixing flow $A(t)$ of a set $A$, its measure of intersection with a fixed
set $B$ tends to the time independent value:%
\begin{equation}
\underset{t\rightarrow \infty }{\lim }\mu (A(t)\cap B)=\mu (A)\mu (B).
\label{Mix}
\end{equation}%
Hence, measuring the amount of liquid $A$ in the fixed set $B$, $\mu
(A(t)\cap B)$, at different times $t$ one will find that after some time the
value of $\mu (A(t)\cap B)$ gets closer and closer to a constant $\mu (A)\mu
(B),$ which can naturally be considered as the equilibrium value. Thus, one
can say that in the mixing flow any nonequilibrium distribution relaxes to
the equilibrium state. The formula (\ref{Mix}) tells one that in a mixing
flow, a liquid fragment $A$ uniformly spreads over the accessible space; As
time elapses, $A$ develops filaments which get thinner and thinner, and its
points decorrelate and move more and more independently. Thus, the mixing
motion of $A$ is chaotic indeed. For instance, if $A$ consists of fragments
of different colors which do not interpenetrate, then after the relaxation
every one of these colors will be present in any fixed set $B$ however small
it be. The only difference to our case is that we have different levels of
the initial DF rather than colors.

The ergodic theory considers flows on a phase space where the state of a
system is given by a single point while a set of finite measure represents
an ensemble of similar systems. The $\mu $ space is not the phase space of a
6D liquid (the phase space is introduced in the next Sec.2). In $\mu $ space
the state of a liquid is determined by a set occupied by the liquid or, in
other words, by liquid's shape. A liquid element does spread over the entire
$\mu $ space, but not necessarily homogeneously as in phase space, and it is
finding the inhomogeneous distribution in the $\mu $ space that is the main
problem of the theory \cite{Ideal gas}. Therefore, in what follows the
homogeneity of smearing expressed by Eq.(\ref{Mix}) is not assumed in the $%
\mu $ space even in the equilibrium state.

Now we can more instructively describe the fundamental problem one has to
face when developing a statistical theory of an actual continuous liquid.
Suppose we know the above distribution of the liquid's levels over the fixed
cells $\{B\}$ in the $\mu $ space. Given these coarse-grained macroscopic
quantities are fixed, however, the arrangement of different levels inside a
single cell can be different at a finer scale. The question is how many
states with different inner arrangements exist for the above fixed
macroscopic levels' distribution. The answer to this question can be
obtained by using the famous Shannon-McMillan-Breiman theorem. The contact
of our problem to this theorem will be established in Sec. III.C.

The statistical property of mixing implies that any small fragment of a
liquid spreads over the total accessible space. This is a complete mixing,
but we would like to make a contact with the more real situation. As already
stressed above, the Vlasov dynamics cannot result in a complete mixing as
the higher the mixing achieved in the system, the smoother the mean field
is, and, as a result, the weaker the mixing efficiency becomes. An
incomplete mixing implies that points of a liquid that were initially
sufficiently remote from one another move independently and can be found at
any relative distance. However, the filaments developing in the flow will
not be going beyond some minimum size, the mixing stops and finer scale
patterns will not be produced. This is because the points of a liquid that
were initially separated insufficiently remain correlated and cannot run
away from one another. Then the degree of incomplete mixing can be accounted
for by the number $M$ of sets in the partition of the initial liquid's shape
$A(0)$ which can be considered statistically independent.

Let $A(0)$ be the initial set occupied by the liquid and $\{A_{m}(0)\},$ $%
m=1,2,...,M$, be its partition into disjoint subsets $A_{m}$, i.e.,
\begin{equation}
A(0)=\bigcup\limits_{m=1}^{M}A_{m}(0).  \label{A0}
\end{equation}%
In the flow, each $A_{m}$ changes its shape, and the argument $0$ in Eq.(\ref%
{A0}) changes to $t$. Let $\{B_{i}\}$ be a partition of the accessible space
into fixed disjoint subsets $B_{i}$. Obviously, the probability that a point
from $A_{m}$ is found in $B_{i}$ is proportional to the quantity $\mu
(A_{m}\cap B_{i}).$ If the liquid elements $A_{m}$ flow independently, then
the joint probability $P$ that a point from $A_{1}$ enters a set $B_{i_{1}}$%
, a point from $A_{2}$ enters a set $B_{i_{2}},...,$ a point from $A_{M}$
enters a set $B_{i_{M}}$ is proportional to the product of the probabilities:%
\begin{equation}
P\{A_{1}\rightarrow B_{i_{1}}|A_{2}\rightarrow
B_{i_{2}}|...|A_{M}\rightarrow B_{i_{M}}\}\propto \prod\limits_{m=1}^{M}\mu
(A_{m}\cap B_{i_{m}}).  \label{PM}
\end{equation}%
It is important to stress that there are only $M$ factors in the joint
probability $P$ as the maximum number of independent events $%
A_{m}(t)\rightarrow B_{i_{m}}$ at a time is $M$. We will see that it is the
length $M$ of a sequence consisting of the above independent events that
determines the number of microscopic states for a given coarse-grained
levels' distribution. As to points within the same set $A_{m}$, by the
assumption of incomplete mixing, they do not move independently and
contribute very little to the number of microscopic states.

I choose the number $M$ of the statistically independent elements of a
liquid as a measure of incomplete mixing. For $M$ very large, i.e., much
larger than the number of particles $N$ in the system, the flow is strongly
mixing. For $M=0$, the system is ergodic without mixing. In the intermediate
case of $M$ on the order of $N$ or lesser, the mixing is incomplete. We will
see that form of the equilibrium DF depends on $M.$

In the next Section, I introduce the phase space of the 6D incompressible
Vlasov liquid, its microscopic and macroscopic states, formulate the problem
in terms of the intersection measures discussed above, and present the exact
constraints on the accessible phase space due to the incompressibility of
the flow. In Sec.III, the microcanonical statistical integral for the
accessible number of states is introduced and formulated in terms of the
macrostates. In particular, in Sections III.B and III.C, analog of the
discrete combinatorics for a continuous systems, which consists of two
steps, is developed: first, I introduce the ensemble of phase subspaces
which accounts for the states of a liquid on all subsets of the phase space;
second, the number of microstates in a given macrostate is found by
connecting the problem with the Shannon-McMillan-Breiman theorem. In
Sec.III.D the energy of the system is expressed in terms of the macroscopic
states. Here the interaction is taken in the form of the Newton or Coulomb
potentials in the 3D space. This restriction is chosen for certainty, but
the method developed in this paper can be applied for any long-range
potential. In Sec.IV, the statistical integral is calculated and in Sec.V
the results are present for strong mixing, for ergodicity without mixing,
and for an incomplete mixing. The end of Sec.III is devoted to some
thermodynamics, in particular, the entropy is expressed in terms of the
total energy and the initial DF. In the conclusion, Sec.VI, the results are
briefly summarized.

The equilibrium DF for strong mixing is found to be exactly of the
Fermi-Dirac form: different levels do not appear individually and are
present as a perfect mixture which is distributed over energy values. A pure
ergodic case without mixing is considered formally since, as explained in
Sec.I.B, in this case the system does not relax to a time independent state.
The formal summation over possible occupation numbers in a small reference
cell results exactly in the LBDF. In the intermediate case of incomplete
mixing, the equilibrium DF is more complex than the LBDF : statistically
independent elements with different initial densities (levels) have
distinguishable contributions in the DF, but, in addition, the contribution
of a given value of the (continuous) occupation number in the $\mu $ space
appears with the weight depending on the mixing degree.

\section{The macroscopic and microscopic states of the Vlasov liquid}

Denote $\Omega $ the 6D $\mu $ space. Let $\tau $ be a point of $\Omega $~: $%
\tau =(\mathbf{r},\mathbf{v})\in \Omega $ where $\mathbf{r}$ is the vector
of position and $\mathbf{v}$ is the velocity of a particle. Let at $t=0$ the
initial DF $f_{0}(\tau _{0})$ be nonzero in the bounded area $\Omega _{0}$
whose points be $\tau _{0}$: $\tau _{0}\in \Omega _{0}\subset \Omega .$
Clearly, the set $\Omega _{0}$ [i.e., support of $f_{0}(\tau _{0})$]
describes the initial shape of the 6D liquid while $f_{0}(\tau _{0})$ gives
the liquid's density at each point $\tau _{0}$. The temporal dynamics of the
liquid in $\Omega $, the flow, is governed by the Vlasov equation. After the
time $t$ each point $\tau _{0}$ moves to the point $\tau (\tau _{0},t)$ and
the liquid's shape changes, $\Omega _{0}\rightarrow \Omega _{0}(t)$. All
liquid elements of $\Omega _{0}$ also change their shapes but the Vlasov
equation preserves their individual volumes which is associated with the
incompressibility. As a result, the value of DF does not change along the
trajectories of points $\tau _{0}:$ $\ $%
\begin{equation}
f_{0}[\tau (\tau _{0},t)]=\ f_{0}(\tau _{0}).  \label{f0}
\end{equation}%
The number of these integrals of motion is uncountable, but they can be
enumerated by the points of the initial state $\Omega _{0},$ one integral
per each $\tau _{0}\in \Omega _{0}.$

The state of a liquid at any instant $t$ is fully determined by the
trajectories of all the initial points $\tau _{0}$, thus the function $\tau
(\tau _{0},t)$ is the liquid's \textit{microstate.} To introduce a
macrostate that can be specified by a countable amount of information, we
resort to the partitions of both moving $\Omega _{0}(t)$ and immobile $%
\Omega $ and introduce the measures of intersections between their members
in line with the ideas outlined in the previous Section.

We devide the $\mu $ space into $i_{m}$ small cells $\sigma _{i}$ of the
volume $\mu (\sigma _{i})=\mu _{i},$ $i=1,2,...i_{m}:$
\begin{equation}
\Omega =\bigcup\limits_{i=1}^{i_{m}}\sigma _{i}.  \label{Omega}
\end{equation}%
These cells are fixed and immobile which is indicated by the subscript $i.$
We also devide the initial set $\Omega _{0}$ into $k_{m}$ small elements $%
\sigma _{0k}$ of the volume $\mu (\sigma _{0k})=\mu _{0k},$ $k=1,2,...k_{m}$%
. The flow $\Omega _{0}\rightarrow \Omega _{0}(t)$ induces the time
evolution of $\sigma _{0k},$ $\sigma _{0k}\rightarrow $ $\sigma _{k}(t),$
which preserves the volume, $\mu (\sigma _{k}(t))=\mu _{0k}.$ The mobile
partition has the form
\begin{equation}
\Omega _{0}(t)=\bigcup\limits_{k=1}^{k_{m}}\sigma _{k}(t).  \label{Omega0}
\end{equation}%
This partition is moving (which is indicated by the subscript $k$,
\textquotedblleft kinesis\textquotedblright , movement) so that each initial
$\sigma _{0k}$ spreads over the $\mu $ space $\Omega $. Spreading over $%
\Omega $, the elements $\sigma _{k}(t)$ intersect the immobile reference
cells $\sigma _{i}.$ Let $\mu _{ik}=\mu (\sigma _{k}\cap \sigma _{i})$ be
the volume of intersection of the spreading liquid element $\sigma _{k}$
with the fixed reference cell $\sigma _{i}.$ The quantity
\begin{equation}
X=\{\mu _{11},\mu _{12,}...,\mu _{21},\mu _{22,}...,\mu _{ik},...,\mu
_{i_{m}k_{m}}\}  \label{X}
\end{equation}%
comprising the measures of all possible intersections is a point of some
space $\Phi $ which has dimension $\dim \Phi =i_{m}k_{m}.$

We assume that $\sigma _{0k}$ is so small that the DF therein is constant: $%
f_{0}(\tau _{0})=f_{k}$ for any $\tau _{0}\in \sigma _{0k}.$ Hence, by
virtue of the incompressibility expressed by the conservation laws (\ref{f0}%
), one has $f_{0}(\tau )=f_{k}$ for any $\tau \in \sigma _{k}(t)$. This
shows that the quantity $X$ gives the coarse-grained distribution of the
levels, which can be labelled by the subscript $k,$ over the reference
cells: $\mu _{ik}$ is the fraction of the liquid with density $f_{k}$ in the
cell $\sigma _{i}.$ The quantity $X$ can therefore be considered as the
\textit{macrostate} and the space $\Phi $ as the phase space of the system.
The integrals of motion given in Eq.(\ref{f0}) can now be expressed as the
following $k_{m}$ constraints in $\Phi :$%
\begin{equation}
\mu _{0k}=\sum_{l=1}^{i_{m}}\text{\ }\mu _{ik},\text{ }k=1,2,...,k_{m}.
\label{mu0k}
\end{equation}%
These constraints express the conservation of the volume of the $k$-th
element of the liquid: this element spreads over all cells $\sigma _{i}$,
but its total amount in all $\sigma _{i}$ is equal to its initial volume $%
\mu _{0k}$. It is more convenient to reformulate the constraints in the form
of matter conservation. To this end we notice that the number of particles
in $\sigma _{k}$ is $N_{k}=f_{k}\mu _{0k}.$ Then multiplying Eq.(\ref{mu0k})
with $f_{k}$ one obtains%
\begin{equation}
N_{k}=\sum_{l=1}^{i_{m}}\text{\ }\mu _{ik}f_{k},\text{ }k=1,2,...,k_{m}.
\label{Nk}
\end{equation}

These constrains do not exhaust all restrictions on the macroscopic states $%
X $. Indeed, the amount of liquid in a cell $\sigma _{i}$ cannot exceed its
volume $\mu _{i}$. Then we have the following $i_{m}$ inequalities:%
\begin{equation}
0\leq \sum_{k=1}^{k_{m}}\text{\ }\mu _{ik}\leq \mu _{i},\text{ }%
i=1,2,...,i_{m}.  \label{sigmai}
\end{equation}

The information about the system depends on the partitions $\{\sigma _{0k}\}$
and $\{\sigma _{i}\}.$ It is known that the information which can be
delivered by a partition is given by the so-called partition entropy \cite%
{Sinai}. The entropies of our partitions are, respectively, $-\sum_{k}\mu
_{0k}\ln \mu _{0k}$ and $-\sum_{i}\mu _{i}\ln \mu _{i}$. Under the
constraints $\sum_{k}\mu _{0k}=\mu (\Omega _{0})$ and $\sum_{i}\mu _{i}=\mu
(\Omega )$ the entropies attain their maxima for $\mu _{0k}=\mu (\Omega
_{0})/k_{m}$ and $\mu _{i}=\mu (\Omega )/i_{m}.$ Thus, the most informative
partitions consist of elements of equal measure. In the following we assume
that $\mu (\Omega _{0})=1$ and choose all the partitions' elements of the
same size $\omega $, i.e.,
\begin{eqnarray}
\mu _{0k} &=&\mu _{i}=1/k_{m}=\omega ,  \label{omega} \\
i &=&1,2,...,i_{m};\text{ }k=1,2,...,k_{m}.  \notag
\end{eqnarray}%
We emphasize however that the shape of the elements $\sigma _{0k}$ and $%
\sigma _{i}$ is not supposed to be necessarily the same. The only
requirement on the shapes is that the partitions are integral ones (so that
for small $\omega $ the summation can be changed to integration). Note that
in spite of their equal values, the symbols $\mu _{0k}$ and $\mu _{i}$ will
often appear in our consideration as their subscripts are helpful for making
formulas clearer and more symmetric. In what follows, making use of the
results of this Section, I formulate the statistical integral and then solve
it by the steepest descent method.

\section{Statistical integral for the 6D incompressible Vlasov liquid}

\subsection{Microcanonical ensemble for the Vlasov liquid}

The equilibrium states of many systems with long-range interaction, e.g.,
galaxies and charged beams, are spatially inhomogeneous, confined by their
own mean field without a contact with any external body. Such systems are
isolated and their statistical equilibrium is microcanonical. Moreover, the
derivation of the canonical ensemble is based on system's additivity and
fixed temperature, that is on two assumption invalid for a system with
long-range interaction. For this reason we consider the microcanonical
statistical integral for an isolated Vlasov system with the energy $E$. In
the context of the previous Section, the number of states $\Gamma $ of our
system is given by the volume of the fraction of the phase space $\Phi $
determined by the $k_{m}$ constraints (\ref{Nk}), $i_{m}$ inequalities (\ref%
{sigmai}), and the requirement of the constant energy. This can be written
as
\begin{equation}
\Gamma =\int d\Sigma \times \delta \lbrack E-H(X)]\times W(\Omega )e^{h(X)},
\label{Gamma}
\end{equation}%
where the integration is over the surface $\Sigma $ in the phase space $\Phi
$ which is determined by the constraints (\ref{Nk}):

\begin{equation}
\int d\Sigma (X)=\prod_{k=1}^{k_{m}}\prod_{i=1}^{i_{m}}\int_{0}^{\omega
}d\mu _{ik}\delta (N_{k}-\sum_{i^{\prime }=1}^{i_{m}}\mu _{i^{\prime
}k}f_{k}).  \label{Sigma}
\end{equation}%
The function $H(X),$ which will be considered in Sec. III.D, is the energy
of the system as a function of its macrostate $X$. The product $W(\Omega
)\exp [h(X)]$ is the cenral problem of the statistical approach to
continuous systems. It is the statistical weight of the state $X$ which is
the continuous analog of the statistical weight well-known for discrete
systems with a countable number of states and finite number of particles.
The factor $W(\Omega )$ determines what I call the ensemble of phase
subspaces, Sec. III.B. It accounts for states with "empty cells" which is
the task more elaborated than that for discrete systems. The quantity $\exp
[h(X)]$ gives the number of microstates in a macrostate $X$ for the given
degree $M$ of incomplete mixing (Sec. III.C). The inequalities (\ref{sigmai}%
) are incorporated in the expression for $W(\Omega )$.

\subsection{Analog of combinatorics for a continuous Vlasov system: The
ensemble of phase subspaces}

Using combinatorics, the number of microstates corresponding to a given
macrostate can be readily computed for a system with a countable number of
states and finite number of particles. The macrostate consists of
macroscopic but small boxes which in turn consist of many microscopic
states. Some of macroscopic boxes can be empty, and the discrete
combinatorics naturally accounts for macroscopic states with different
occupation numbers in some boxes while keeping a number of other boxes
empty. We need a generalization of this counting to a continuous system with
uncountable number of states.

As outlined above, amount of liquid in a fixed macroscopic cell $\sigma _{i}$%
, which plays the role of the occupation number, is a continuous variable
ranging between $0$ and $\omega $. Consequently, the summation over the
occupation numbers is replaced by the integration over $d\Sigma (X)$, Eq.(%
\ref{Sigma}), which is the measure in the space $\Phi $ of dimension $%
i_{m}k_{m}.$ Now the states with, say, $l$ "empty boxes" are those with $\mu
_{ik}$ kept zero for some $l$ subscripts $i$ from the set $\{i\},$ \ while
for the other $(i_{m}-l)$ subscripts $i^{\prime }$ the variables $\mu
_{i^{\prime }k}$ are varying from $0$ to $\omega $. But such states reside
on the subspaces $\Phi _{(i_{m}-l)}$ of $\ $the $\dim $ension $%
(i_{m}-l)k_{m}<\dim \Phi $ , hence their measure in the space $\Phi $ is
zero (e.g., 2D surfaces and 1D lines in a 3D volume have measures $0$). To
account for the possible realizations of a continuous system on the
subspaces of $\Phi $, one has to insert the singular density $%
\prod\nolimits_{k}2\delta (\mu _{ik})$ for any empty cell $\sigma _{i}$ ($%
\int_{0}^{\omega }d\mu _{ik}\delta (\mu _{ik})=1/2$ and the coefficient $2$
makes it $1$). As $\{\sigma _{i}\}$ is the partition of the $\mu $ space $%
\Omega $, summation over all possible distributions of empty cells $\sigma
_{i}$ is equivalent to a summation over all subspaces of $\Omega $, which is
indicated by the argument of the singular density $W$ in the integral $%
\Gamma .$

Let $\{i_{1},i_{2},...,i_{l}\}$ be a subset of $\{i\}$ consisting of some $l$
possible values of $i$. Then summing over all possible distributions of
empty cells $\sigma _{i}$ implies the following form of $W:$%
\begin{equation}
W(\Omega )=\sum_{l=0}^{i_{m}}\sum_{i\in \{i_{1},i_{2},...,i_{l}\}}\delta
\{i_{1},i_{2},...,i_{l}\},  \label{W}
\end{equation}%
where%
\begin{equation}
\delta \{i_{1},i_{2},...,i_{l}\}\text{\ }=\prod_{i\in
\{i_{1},i_{2},...,i_{l}\}}\prod_{k=1}^{k_{m}}2\delta (\mu _{ik})\times
\prod_{i^{\prime }\notin \{i_{1},i_{2},...,i_{l}\}}\Theta \left( \mu
_{i^{\prime }}-\sum_{k^{\prime }=1}^{k_{m}}\mu _{i^{\prime }k^{\prime
}}\right) .  \label{W1}
\end{equation}%
The density $\delta \{i_{1},i_{2},...,i_{l}\}$ is singular on the single
subspace of $\Omega $ consisting of $l$ empty cells $\sigma _{i}$ with $i\in
\{i_{1},i_{2},...,i_{l}\}.$ The inner summation in (\ref{W}) is over all
possible subspaces of $\Omega $ consisting of $l$ cells, and the second sum
is over all possible $l$ $.$ The structure of the singular density $\delta
\{i_{1},i_{2},...,i_{l}\}$ is as follows: each empty cell $\sigma _{i},i\in
\{i_{1},i_{2},...,i_{l}\},$ has the delta functions $\delta (\mu _{ik})$ for
all $k$, and each nonempty cell $\sigma _{i^{\prime }},$ $i^{\prime }\notin
\{i_{1},i_{2},...,i_{l}\},$ has the Heaviside step function $\Theta $ which
accounts for the inequality (\ref{sigmai}). $\ $We see that $W(\Omega )$ is
the singular density picking up all possible distributions of empty cells
over the $\mu $ space $\Omega .$ Equivalently, $W(\Omega )$ provides the
contributions from the arrangements of the system on all subspaces of the
phase space $\Phi $ and represents the ensemble of phase subspaces. Of
course, the liquid of volume $\mu (\Omega _{0})$ cannot be inserted into a
subspace of the volume lesser than $\mu (\Omega _{0})$ as the appropriate
states require minimum $\mu (\Omega _{0})/\omega $ cells for its
arrangement. But the inappropriate states on insufficient subspaces will be
eliminated due to the constrains (\ref{mu0k}). This is similar to the use of
the grand canonical ensemble for a system of $N$ particles. This ensemble
counts for the contributions from any number of particles and then the spare
contributions are eliminated by the appropriate choice of the chemical
potential. The analogy is even deeper. The grand partition function
factorizes because as the restriction on $N$ is removed, the states of the
systems do not correlate. In the ensemble of phase subspaces, there is no
restriction on the liquid's size and, as a result, the density $W(\Omega )$
factorizes over the cells $\sigma _{i}$. It is not difficult to realize that
\begin{equation}
W(\Omega )=\prod_{i=1}^{i_{m}}\left[ \prod_{k=1}^{k_{m}}2\delta (\mu
_{ik})+\Theta \left( \mu _{i^{\prime }}-\sum_{k^{\prime }=1}^{k_{m}}\mu
_{i^{\prime }k^{\prime }}\right) \right] .  \label{W2}
\end{equation}%
We will also need the expression of the form $W(\Omega )\exp (\sum_{i}a_{i})$
where $a_{i}=0$ for $\mu _{ik}=0.$ This expression also factorizes to give%
\begin{equation}
W(\Omega )\exp \left( \sum_{i}a_{i}\right) =\prod_{i=1}^{i_{m}}\left[
\prod_{k=1}^{k_{m}}2\delta (\mu _{ik})+\Theta \left( \mu _{i^{\prime
}}-\sum_{k^{\prime }=1}^{k_{m}}\mu _{i^{\prime }k^{\prime }}\right) e^{a_{i}}%
\right] .  \label{Wa}
\end{equation}%
It is impotant to stress that it is the accounting for empty cells $\sigma
_{i}$ that results in the Fermi-like \ properties of the equilibrium states
found below: without the delta functions in (\ref{W2}) the resulting DFs can
only be Maxwell-Boltzmann-like.

The developed ensemble of phase subspaces solves the problem of picking up
the states with empty macroscopic "boxes" for the continuous systems, but
this is not yet the full analog of the discrete combinatorics. Now we need
to find the number of microscopic states $\exp [h(X)]$ in a given
macroscopic state $X$. For continuous systems this problem can be addressed
using the ideas and results of the ergodic theory.

\subsection{The Shannon-McMillan-Breiman theorem and the number of
microstates in a given macrostate\textit{\ }$X$}

Assume that the incomplete mixing of the Vlasov flow is described by the
parameter $M$ introduced in Sec.I.B. This means that there is a partition of
the liquid body into $M$ elements which move independently. Then each mobile
liquid element $\sigma _{0k}$ contains $M/k_{m}=M_{k}$ independent
subelements $\sigma _{k,m}$, $m=1,2,...,M_{k}$ (we assume that $M_{k}$ does
not depend on $k$, but keep the subscript $k$ for clarity). In order to
apply a statistical approach, we have to asume that a reference cell can
have contributions from many different $\sigma _{k,m}$ with the same $k$.
Thus we assume that
\begin{equation}
M\gg M/k_{m}=M_{k}\gg i_{m}\geq k_{m}.  \label{MIK}
\end{equation}

Consider first subelements $\sigma _{k,m}$ of the mobile element with a
fixed subscript $k$. In the flow, $\sigma _{k,m}$ spread over the reference
cells $\sigma _{i},$ resulting in nonzero intersections $\mu _{ik,m}=\mu
(\sigma _{k,m}\cap \sigma _{i}).$ The quantity $\mu _{ik,m}$ is proportional
to the probability $p_{i,km}$ that a point from the subelement $\sigma
_{k,m} $ can be found in the reference cell $\sigma _{i}$. As all the
subelements in the mobile $\sigma _{k}$ have the same density $f_{k},$ the
probability $p_{i,km}$ is also the same for all the subelements and does not
depend on $m$, i.e., $p_{i,km}=p_{ik}$. On the other hand, $\mu
_{ik}=\sum\nolimits_{m}\mu _{i,km},$ hence $p_{ik}\propto \mu _{ik}/M_{k}$.
To find the proportionality coefficient, we reqire that
\begin{equation}
\sum_{i=1}^{i_{m}}p_{ik}=1  \label{p}
\end{equation}%
and recall the relations (\ref{mu0k}) and (\ref{omega}), which gives
\begin{equation}
p_{ik}=\mu _{ik}/\omega .  \label{pik}
\end{equation}%
Let us choose one point from each subelement: $\tau _{k,m}\in \sigma _{k,m}$%
, $m=1,2,...,M_{k}$. Then, in line with the idea presented in Sec.I.B, the
probability of a joint event that $\tau _{k,1}$ enters $\sigma _{i_{1}},$ $%
\tau _{k,2}$ enters $\sigma _{i_{2}},...,$ $\tau _{k,M_{k}}$ enters $\sigma
_{i_{M_{k}}},$ is the following product:%
\begin{equation}
P\{\tau _{k,1}\rightarrow \sigma _{i_{1}}|\tau _{k,2}\rightarrow \sigma
_{i_{2}}|...|\tau _{k,M_{k}}\rightarrow \sigma
_{i_{M_{k}}}\}=\prod\limits_{m=1}^{M_{k}}p_{ik}.  \label{P}
\end{equation}%
Now we formalize the above properties of the $M_{k}$ subelements of a mobile
element $\sigma _{k}$ spreading over the $\mu $ space $\Omega $. We have a
finite set of states $\{\sigma _{i}\}$ consisting of $i_{m}$ elements $%
\sigma _{i}$, and the probabilistic measure $p_{ik}$ on $\{\sigma _{i}\}$: $%
p(\sigma _{i})=p_{ik}.$ The argument of the above joint probability $P$ is a
sequence $\mathbf{\omega }_{_{k}}$ which can be briefly presented as $%
\mathbf{\omega }_{_{k}}=\{\sigma _{i_{1}}(\mathbf{\mathbf{\omega }}_{_{k}}%
\mathbf{)},$ $\sigma _{i_{2}}(\mathbf{\omega }_{_{k}}\mathbf{),}...,\sigma
_{i_{M_{k}}}(\mathbf{\omega }_{_{k}}\mathbf{)\}.}$ One can say that this is
a possible realization of a large experiment $\mathbf{\omega }_{_{k}}$
consisting of $M_{k}$ simple experiments: output of the first simple
experiment is $\sigma _{i_{1}}(\mathbf{\mathbf{\omega }_{_{k}}),}$ of the
second is $\sigma _{i_{2}}(\mathbf{\mathbf{\omega }}_{_{k}}\mathbf{),}$ ...,
and, finally, output of the last simple experiments is $\sigma _{i_{M_{k}}}(%
\mathbf{\omega }_{_{k}}\mathbf{).}$ The outputs of different simple
experiments are independent so that the probability of the large experiment $%
\mathbf{\mathbf{\omega }}_{_{k}}$ is given by the product
\begin{equation}
P\{\mathbf{\mathbf{\omega }}_{_{k}}\}=\prod\limits_{m=1}^{M_{k}}p_{ik}.
\label{PP}
\end{equation}%
The total number of different observable sequences $\mathbf{\mathbf{\omega }}%
_{_{k}}$ (i.e., with nonzero $P\{\mathbf{\mathbf{\omega }}_{_{k}}\}$) gives
the desired number of the microscopic states for a macroscopic state $X$ (%
\ref{X}) by virtue of the relation (\ref{pik}).

The obtained stochastic process is exactly the well-known Bernulli map and
the sequence $\mathbf{\mathbf{\omega }}_{_{k}}$ of an independent random
variables with the values from $\{\sigma _{i}\}$ is an element of its phase
space \cite{Bil,Sinai}. Strictly speaking, the phase space of Bernulli map
consists of infinite sequences, but this is not an obstacle for our task of
finding the total number of different observable sequences of a large length
$M_{k}.$ This number is given by the Shennon-McMillan-Breiman theorem \cite%
{Bil,Sinai} which, for our purpose, can be formulated as follows: for large $%
M_{k}$, the probability $P\{\mathbf{\mathbf{\omega }}_{_{k}}\}$ of any
observable sequence $\mathbf{\mathbf{\omega }}_{_{k}}$ is close to $\exp
(-M_{k}h_{k})$ where $h_{k}$ is the Kolmogorov-Sinai entropy of a given
Bernulli map. The Kolmogorov-Sinai entropy is determined by the probability
measures $p_{ik}$ on the space of state $\{\sigma _{i}\}$ and is given by
the following formula:%
\begin{equation}
h_{k}=-\sum_{i=1}^{i_{m}}p_{ik}\ln p_{ik}.  \label{hk}
\end{equation}%
The total number of different observable sequences is equal to $1/P$. Thus,
the number of microstates $g_{k}$ corresponding to the same macrostate $X$ (%
\ref{X}) with a fixed $k$ is
\begin{equation}
g_{k}=\exp \left( -M_{k}\sum_{i=1}^{i_{m}}p_{ik}\ln p_{ik}\right)  \label{gk}
\end{equation}%
As subelements of the mobile elements $\sigma _{k}$ with different $k$ are
independent, the total number of microstates corresponding to a given
macrostate $X$ with any $k$ is the product $\prod\nolimits_{k}g_{k}.$ Then
the quantity $h(X)$ in the statistical integral $\Gamma $ (\ref{Gamma}) is
finally obtained in the form%
\begin{equation}
h(X)=-M\sum_{i=1}^{i_{m}}\sum_{k=1}^{k_{m}}\mu _{ik}\ln (\mu _{ik}/\omega ).
\label{h}
\end{equation}

Now we know the $X$ dependence of all of the quantities in the integral $%
\Gamma $ (\ref{Gamma}) but the energy $H$. The function $H(X)$ is considered
below.

\subsection{Energy as a function of the macroscopic state $X$}

For further consideration it is convenient to use a simplified notation for
the coordinates $\tau $ in the $\mu $ space. A point $\tau $ from the
reference element $\sigma _{i}$ will be denoted as $\tau _{i}$ or just $i:$ $%
\tau _{i}\equiv i=(\mathbf{r}_{i},\mathbf{v}_{i}).$ The 6D volume $\omega $
of $\sigma _{i}$ is a product $\omega =\omega _{r}\omega _{v}$ where $\omega
_{r}$ and $\omega _{v}$ are its 3D volumes in the $\mathbf{r}$ and $\mathbf{v%
}$ spaces. The number of particles $n_{i}$ in a fixed element $\sigma _{i}$
of the $\mu $ space is
\begin{equation}
n_{i}=\sum_{k=1}^{k_{m}}\mu _{ik}f_{k}.  \label{ni}
\end{equation}%
We aslo need the number of particles $n_{r_{i}}$ in the cell $\omega
_{r_{i}} $ of the space of coordinates :%
\begin{equation}
n_{r_{i}}=\sum_{v_{i}}n_{i},  \label{nr}
\end{equation}%
In the mean field approximation, the particle energy depends only on the
macroscopic state $X$. Then the total energy $H$ of the system can be
written in the form%
\begin{equation}
H(X)=\sum_{i}\text{\ }\varepsilon _{ad,i}n_{i}+\frac{q^{2}}{2}%
\sum_{i,i^{\prime }}w_{ii^{\prime }}n_{r_{i}}n_{r_{i^{\prime }}},  \label{H}
\end{equation}%
where $\varepsilon _{ad,i}=m\mathbf{v}_{i}^{2}/2+q\varphi _{ext}(\mathbf{r}%
_{i})$ is the additive particle energy and $w_{ii^{\prime }}=w(\mathbf{r}%
_{i}-\mathbf{r}_{i^{\prime }})$ is the pair interaction potential. The
function $q\varphi _{ext}(\mathbf{r}_{i})$ is the particle energy in the
external potential $\varphi _{ext}$ and $q$ is the appropriate constant. For
instance, $q$ is the charge for charged particles and $q$ is the particle's
mass $m$ for gravitating particles. For certainty, we consider only a
Coulomb-like potential. In the 3D geometry, the potential $w_{ii^{\prime }}$
has the form%
\begin{equation}
w_{ii^{\prime }}=-\frac{\gamma }{\left\vert \mathbf{r}_{i}-\mathbf{r}%
_{i^{\prime }}\right\vert },  \label{w}
\end{equation}%
where $\gamma >0$ is Newton's constant for gravitational systems, and $%
\gamma =-1$ for Coulomb interaction. \ Now the integrand of $\Gamma $ is
defined as a function of $\mu _{ik}$ and it remains to perform the
integration.

\section{Solving the statistical integral $\Gamma $}

\subsection{Explicite expression for $\Gamma $}

In the formulas which follow we omit insignificant factors. We use the
following \ analytical representation of the delta functions entering $%
\Gamma $ (\ref{Gamma}):%
\begin{eqnarray}
\delta (E-H(X)) &=&\int\limits_{-\infty }^{\infty }d\beta e^{i\beta
(E-H(X))},  \label{delH} \\
\delta \left( N_{k}-\sum_{i}\mu _{ik}f_{k}\right) &=&\int\limits_{-\infty
}^{\infty }d\zeta _{k}\exp i\zeta _{k}\left( N_{k}-\sum_{i}\mu
_{ik}f_{k}\right) .  \label{delNk}
\end{eqnarray}%
For the Heaviside teta function it is convenient to use the following
identity:
\begin{equation}
\Theta \left( \omega -\sum_{k}\mu _{i^{\prime }k^{\prime }}\right)
=\int_{0}^{\omega }d\nu _{i}\delta \left( \nu _{i}-\sum_{k}\mu _{i^{\prime
}k^{\prime }}\right) .  \label{Teta}
\end{equation}%
To get rid of the terms quadratic in the spatial occupation numbers $%
n_{r_{i}}$ in the function $\exp (-i\beta H(X)),$ we employ the
Habbard-Stratonovich transformation in the following form \cite{Complex
Gauss}:%
\begin{eqnarray}
&&\exp \left( -\frac{i\beta q^{2}}{2}\sum_{i,i^{\prime }}w_{ii^{\prime
}}n_{r_{i}}n_{r_{i^{\prime }}}\right)  \label{HS} \\
&\propto &(i\beta )^{-i_{m}/2}\prod_{r_{i}}\int\limits_{-\infty }^{\infty
}d\psi _{r_{i}}\exp \left( q\sum_{i}n_{r_{i}}\psi _{r_{i}}+\frac{1}{2i\beta }%
\sum_{i,i^{\prime }}w_{ii^{\prime }}^{-1}\psi _{r_{i}}\psi _{r_{i^{\prime
}}}\right) ,  \notag
\end{eqnarray}%
where $\psi _{r_{i}}$ is a field defined at each spatial point $\mathbf{r}%
_{i}$ and $w_{ii^{\prime }}^{-1}$ is the inverse to the matrix (\ref{w}). We
will need the operator $w_{ii^{\prime }}^{-1}$ in the continuous limit $%
\omega _{r}\rightarrow 0.$ In this limit the inverse operator $w^{-1}$ is
defined as
\begin{equation*}
\int d\mathbf{r}^{\prime \prime }w^{-1}(\mathbf{r}-\mathbf{r}^{\prime \prime
})w(\mathbf{r}^{\prime }-\mathbf{r}^{\prime \prime })=\delta (\mathbf{r}-%
\mathbf{r}^{\prime })
\end{equation*}%
and is known to be
\begin{equation}
w^{-1}(\mathbf{r}-\mathbf{r}^{\prime })=\frac{1}{4\pi \gamma }\delta (%
\mathbf{r}-\mathbf{r}^{\prime })\triangle _{\mathbf{r}^{\prime }},
\label{w-1}
\end{equation}%
where $\triangle _{\mathbf{r}^{\prime }}$ is Laplace's operator with respect
to $\mathbf{r}^{\prime }.$ In the continuous limit the matrix $w_{ii^{\prime
}}^{-1}$ in (\ref{HS}) goes over into $\omega _{r}^{6}w^{-1}(\mathbf{r}_{i}-%
\mathbf{r}_{i^{\prime }}).$

For brevity, we use the notations $D\zeta _{k}=\prod\nolimits_{k}d\zeta _{k}$
and $D\psi _{r_{i}}=\prod\nolimits_{i}d\psi _{r_{i}}$. Then, making use of
the above results of this Section and formulas (\ref{Wa}) and (\ref{h}), the
statistical integral (\ref{Gamma}) can be reduced to the following form:%
\begin{eqnarray}
\Gamma &=&\int\limits_{-\infty }^{\infty }d\beta \beta
^{-i_{m}/2}\int\limits_{-\infty }^{\infty }D\zeta _{k}\int\limits_{-\infty
}^{\infty }D\psi _{r_{i}}\exp \left( i\beta E+i\zeta _{k}N_{k}+\frac{1}{8\pi
\gamma i\beta }\int d\mathbf{r}\psi \triangle \psi \right)  \label{Gamma1} \\
&&\times \exp \sum_{i}\ln (1+Z_{i}).  \notag
\end{eqnarray}%
Here $Z_{i}$ is a fuctional of $\mu _{ik}$ and $\nu _{i}$:
\begin{equation}
Z_{i}=(1/\omega )\int\limits_{0}^{\omega }d\nu _{i}\left(
\prod_{_{k}}\int\limits_{0}^{\omega }d\mu _{ik}\right) \delta \left( \nu
_{i}-\sum\limits_{k^{\prime }}\mu _{ik^{\prime }}\right) e^{a_{i}(\alpha
_{ik})},  \label{Z}
\end{equation}%
where
\begin{eqnarray}
a_{i} &=&\sum_{k=1}^{k_{m}}\left[ \alpha _{ik}\mu _{ik}f_{k}-M\mu _{ik}\ln
(\mu _{ik}/\omega )\right] ,  \label{a} \\
\alpha _{ik} &=&q\psi _{r_{i}}-i\beta \varepsilon _{ad,i}-i\zeta _{k}.
\label{alfa}
\end{eqnarray}%
Now we calculate $Z_{i}$.

\subsection{ The function $Z_{i}$ for different degrees of mixing}

As $M$ and $\mu _{ik}f_{k}$ are large numbers, we can perform the $\mu $
integration using the steepest descent method. Eqs.(\ref{Z}) and (\ref{a})
show that the result is determined by the maximum of $a_{i}$ with respect to
$\mu _{ik}$ under the condition that $\sum\nolimits_{k}\mu _{ik}=\nu _{i}$
for each $i$. This extremum problem is solved by%
\begin{equation}
\overline{\mu }_{ik}=\frac{\nu _{i}\exp (\alpha _{ik}N_{k}/M_{k})}{z_{i}},
\label{mu0}
\end{equation}%
where $N_{k}=\omega f_{k}$, $M_{k}=M/k_{m}$, and $z_{i}$ is the local
partition function
\begin{equation}
z_{i}=\sum\limits_{k=1}^{k_{m}}\exp (\alpha _{i}N_{k}/M_{k}).  \label{z}
\end{equation}%
Changing to $x=\nu _{i}k_{m}$, the result of the $\mu $ integration is
\begin{equation}
Z_{i}=\int\limits_{0}^{1}dx\exp \left[ a_{i}(\overline{\mu }_{ik})+b_{i}%
\right] ,  \label{Z1}
\end{equation}%
where
\begin{subequations}
\begin{equation}
b_{i}=\ln \sqrt{\frac{(2\pi )^{k_{m}}}{\left\vert \det \left\Vert \frac{%
\partial ^{2}a_{i}}{\partial \mu _{ik}\partial \mu _{ik^{\prime }}}%
\right\Vert \right\vert }}  \label{b}
\end{equation}%
The simple calculation gives
\end{subequations}
\begin{equation}
a_{i}(\overline{\mu }_{ik})=M_{k}x\ln (z_{i}/x),  \label{am}
\end{equation}%
$\frac{\partial ^{2}a_{i}}{\partial \mu _{ik}\partial \mu _{ik^{\prime }}}%
=-\delta _{kk^{\prime }}/(\overline{\mu }_{ik}\omega )$, and
\begin{equation}
b_{i}=-\frac{k_{m}}{2}\left[ \ln (Mk_{m}/2\pi )+\ln (z_{i}/x)\right]
+\sum\limits_{k=1}^{k_{m}}\frac{\alpha _{ik}N_{k}}{M_{k}}.  \label{b1}
\end{equation}

In order to proceed, we have to make an assumption concerning the relative
values of the number $M_{k}$ of independent subelements and the number of
particles $N_{k}$ in a mobile element $\sigma _{k}$. Clearly, if $M_{k}\gg
N_{k}$, the mixing is strong whereas if $M_{k}\ll N_{k}$ the mixing is
practically absent and the system is at best ergodic. The intermediate case $%
M_{k}\sim N_{k}$ corresponds to an incomplete mixing. We will consider these
cases separately. Note that the above steepest descent method can be used
only if $M_{k}$ is large, hence the ergodic case without mixing, $M\sim 0$,
has to be considered individually, Sec. V.B.

We first proceed with the case $M_{k}\gg N_{k}$ and $M_{k}\sim N_{k}$. The
inequality (\ref{MIK}) shows that the last two terms in $b_{i}$ can be
neglected compared to $a_{i}$ so that $b_{i}\simeq $ $-\frac{k_{m}}{2}\ln
(M/2\pi \omega )$. This constant can be absorbed into $\zeta _{k}$ [see Eqs.(%
\ref{a}) and (\ref{alfa})], and (\ref{Z1}) reduces to the form
\begin{equation}
Z_{i}=\int\limits_{0}^{1}dxe^{a_{i}(x)}.  \label{ZZ}
\end{equation}

\textit{Strong mixing.} Consider first the case of strong mixing when $%
M_{k}\gg N_{k}$. In this case $a_{i}$ (\ref{am}) can be computed using the
small parameter $N_{k}/M_{k}$. First we consider the function $z_{i}^{M_{k}}$
which is present in $a_{i}$:%
\begin{eqnarray}
z_{i}^{M_{k}} &=&\left[ \sum\limits_{k=1}^{k_{m}}\left( 1+\frac{\alpha
_{ik}N_{k}}{M_{k}}\right) \right] ^{M_{k}}  \notag \\
&=&k_{m}^{M_{k}}\left( 1+\sum\limits_{k=1}^{k_{m}}\frac{\alpha _{ik}N_{k}}{%
M_{k}k_{m}}\right) ^{M_{k}}  \label{exp} \\
&&\underset{M_{k}\rightarrow \infty }{\rightarrow }k_{m}^{M_{k}}\exp
\left\langle \alpha N\right\rangle _{i},  \notag
\end{eqnarray}%
where
\begin{equation}
\left\langle \alpha N\right\rangle _{i}=\sum\limits_{k=1}^{k_{m}}\frac{%
\alpha _{ik}N_{k}}{k_{m}}  \label{alfN}
\end{equation}%
is the average of $\alpha _{ik}N_{k}$ over all $k$. Substituting this to $%
a_{i}$, one has%
\begin{equation}
Z_{i}=\int\limits_{0}^{1}dx\exp [\left\langle \alpha N\right\rangle
_{i}x-M_{k}x\ln (x/k_{m})].  \label{Z11}
\end{equation}%
The exponent in the above integral does not have a maximum inside the
integration interval, but at its right end $x=1$ the exponent has a large
positive derivative and the maximum value. Then, neglecting terms $\sim $ $%
N_{k}/M_{k},$ one obtains:%
\begin{equation}
Z_{i}=\exp (\left\langle \alpha N\right\rangle _{i}+c),  \label{Zstrong}
\end{equation}%
where $c=\ln (k_{m}^{M_{k}}/M_{k}\ln k_{m})$ is a constant.

\textit{Incomplete mixing.} For $M_{k}\sim N_{k}$, the simplification
resulting in the expression (\ref{exp}) does not apply, and one has to deal
with the local partition function in the form (\ref{z}). Then one has:%
\begin{equation}
Z_{i}=\int_{0}^{1}dxz_{i}^{M_{k}x}x^{-M_{k}x}.  \label{ZM}
\end{equation}

The expressions (\ref{Zstrong}) and (\ref{ZM}) for $Z_{i}$ obtained for
different degrees of mixing are essential for the final results. Now we
first present the result for strong mixing, then for the ergodic case
without mixing, and finally for the intermediate mixing.

\section{The equilibrium states of a Vlasov liquid}

\subsection{Maximum entropy state for strong mixing}

In the statistical integral (\ref{Gamma1}) we integrate by part the term $%
\psi \triangle \psi $, replace the summation over $i$ by integration over $%
d\tau =d\mathbf{r}d\mathbf{v}$, and substitute expression (\ref{Zstrong})
for $Z_{i}.$ The result is

\begin{equation}
\Gamma =\int\limits_{-\infty }^{\infty }d\beta \beta
^{-i_{m}/2}\int\limits_{-\infty }^{\infty }D\zeta _{k}\int\limits_{-\infty
}^{\infty }D\psi _{r_{i}}\exp \widetilde{S},  \label{Gamma2}
\end{equation}%
where
\begin{equation}
\widetilde{S}=i\beta E+\sum_{k}i\zeta _{k}N_{k}+\int d\mathbf{r}\left[ -%
\frac{(\mathbf{\nabla }\psi )^{2}}{8\pi \gamma i\beta }+\int \frac{d\mathbf{v%
}}{\omega }\ln (1+e^{\left\langle \alpha N\right\rangle +c})\right] .
\label{S1}
\end{equation}%
This integral can be solved by the steepest descent method because the terms
of $\widetilde{S}$ are proportional to the large total number of particles $%
N $ (we shall see that the first term in the bracket is proportional to the
total interaction energy). In view of the value of $i\beta $ at the saddle
point which is expected to be real, the integration order in (\ref{Gamma2})
is essential. The $\psi $ integral must be performed first\ for imaginary $%
i\beta $ as then it converges both for positive and negative $\gamma $. Then
the integration contours in $\beta $ and $\zeta $ can be deformed down as to
cross the imaginary axes at negative values.

The extremum configuration which gives the dominating contribution to $%
\Gamma $ is determined by the three equations:%
\begin{equation}
\frac{\delta \widetilde{S}}{\delta \psi }=\frac{\partial \widetilde{S}}{%
\partial i\beta }=\frac{\partial \widetilde{S}}{\partial (i\zeta _{k}N_{k})}%
=0.  \label{delS}
\end{equation}%
It is straightforward to realize that the third equation for the quantity $%
\zeta _{k}N_{k}$ is exactly the same for all $k$, hence $\zeta _{k}N_{k}$
does not depend on $k$. Then the best way to present and consider the above
three equations is to change to the following variables:%
\begin{eqnarray}
i\beta &=&\frac{1}{T},\text{ }  \notag \\
\psi &=&-i\beta \varphi ,  \label{T mu fi} \\
-i\zeta _{k}+c/\overline{n} &=&\frac{\eta }{T},  \notag
\end{eqnarray}%
where $\overline{n}=N/k_{m}$ is the average number of particle per a mobile
element $\sigma _{k}.$ We also introduce the average density of the liquid $%
\overline{f}_{0}=N/\mu (\Omega _{0})$ where $\mu (\Omega _{0})$ is its 6D
volume in the natural units, and the total particle energy \ $\varepsilon
=\varepsilon _{ad}+q\varphi :$
\begin{equation*}
\varepsilon =\frac{m\mathbf{v}^{2}}{2}+q(\varphi _{ext}+\varphi ).
\end{equation*}%
Making use of Eqs.(\ref{T mu fi}) and (\ref{alfN}) and $\varepsilon $, one
has
\begin{equation}
\left\langle \alpha N\right\rangle _{i}+c=\frac{\eta -\varepsilon _{i}}{T}%
\overline{n}.  \label{alfN1}
\end{equation}%
Now the equations (\ref{delS}) can be reduced to the following system:
\begin{eqnarray}
\triangle \varphi &=&4\pi \gamma q\int d\mathbf{v}\overline{f},  \label{fi}
\\
E &=&\int d\mathbf{r}\left[ -\frac{(\mathbf{\nabla }\varphi )^{2}}{8\pi
\gamma }+\int d\mathbf{v}\overline{f}\varepsilon \right] ,  \label{E} \\
\int d\mathbf{v}d\mathbf{r}\overline{f} &=&N.  \label{N}
\end{eqnarray}%
where%
\begin{equation}
\overline{f}(\mathbf{r},\mathbf{v})=\overline{f}_{0}\frac{\exp \frac{\eta
-\varepsilon (\mathbf{r},\mathbf{v})}{T}\overline{n}}{1+\exp \frac{\eta
-\varepsilon (\mathbf{r},\mathbf{v})}{T}\overline{n}}.  \label{DF1}
\end{equation}%
The above equations have a clear interpretation. Eq.(\ref{fi}) is the
Poisson equation for the mean field potential $\varphi $. Eq.(\ref{E})
expresses the fact that the energy of the system is equal to $E$. Eq.(\ref{N}%
) shows that the function $\overline{f}$ is normalized on the total number
of particles $N$. Finally, this function $\overline{f}$ is the equilibrium
DF in the $\mu $ space. For $\gamma $ being Newton's constant and $q$ being
mass of a particle, the system is gravitational; While for $\gamma =-1$ and $%
q$ being the electrostatic charge, this is a Coulomb system.

We see that the DF is exactly of the Fermi-Dirac form, different levels $%
f_{k}$ of the initial DF appear only in the mixture $\overline{n}%
=\sum\nolimits_{k}f_{k}\omega /k_{m}=N/k_{m}.$ This is in contrast to the
LBDF (see Eq.(\ref{LBDF}) below) which is the sum of contributions from
different levels $f_{k}$ each with its own chemical potential and
distribution over the energy values. The obtained result implies that in the
case of strong mixing, there are two relaxation processes, fast and slow.
The fast consists of creation and maintaining of a perfect mixture with the
density $N/\mu (\Omega _{0}),$ and the second, slow process consists of
distributing this mixture over the possible energies without changing the
local composition of different levels. The fast relaxation is driven by the
fine scale gradients of the mean field which at this scale must be violent
indeed. The slow relaxation is driven by large scale gradients. This is a
highly idealized relaxation with strong mixing, but if it is assumed, then
the DF is given by Eq.(\ref{DF1}).

\subsection{Ergodicity without mixing and its relation to the LBDF}

Now we consider the result under the assumption that the flow is ergodic
without mixing, i.e., $M=0$. As was explained in Sec.I.B, an ergodic
nonmixing system does not tend to a time independent state. Therefore, all
we can do is to find the number of all states through which the system will
pass over a very large time. In absence of mixing the shape of area $\Omega
_{0}(0)$ occupied by liquid changes only on a large scale, the small moving
elements $\sigma _{k}$ do not spread, fine scale filaments are not produced,
and the boundary of $\Omega _{0}(t)$ does not develop into a long and
intricate line. This means that, with a high accuracy, the occupation number
in a reference cell $\sigma _{i},$ when it is traversed by a $k$-th mobile
element $\sigma _{k},$ is $f_{k}\omega .$ Thus, the occupation number in any
reference cell runs over the values $f_{k}\omega ,$ $k=1,2,...k_{m}.$ Hence,
summation over these numbers must replace the $\mu $\ and $\nu $ integration
in the formula for $Z_{i}$ (\ref{Z}) \cite{Phys Lett}. This gives:%
\begin{equation}
Z_{i}=\sum_{k=1}^{k_{m}}\exp (\alpha _{ik}f_{k}\omega ).  \label{Zerg}
\end{equation}%
Now $\zeta _{k}$ depends on $k$ so that instead of third line in Eq.(\ref%
{delS}) we put $-i\zeta _{k}=\frac{\eta _{k}}{T}.$ Now the sum over $k$ is
transformed to the integral over the initial area $\Omega _{0}$ of the $\mu $
space occupied by the liquid and the subscript $k$ is changed to the
argument $\tau _{0}\in \sigma _{k}$. Then $\eta _{k}+const\rightarrow \eta
(\tau _{0})$ and $Z_{i}$ (\ref{Zerg}) is replaced by the following integral%
\begin{equation}
Z(\mathbf{r},\mathbf{v})=\int_{\Omega _{0}}d\tau _{0}\exp \left[ \frac{\eta
(\tau _{0})-\varepsilon (\mathbf{r},\mathbf{v})}{T}f_{0}(\tau _{0})\omega %
\right] .  \label{Zerg1}
\end{equation}%
Making use of this $Z_{i}$ in $\widetilde{S}$ (\ref{S1}) instead of $%
e^{\left\langle \alpha N\right\rangle +c}$ and solving the statistical
integral (\ref{Gamma1}) as before, one arrives at the following system of
equation. The equations $\delta \widetilde{S}/\delta \psi =0$ and $\partial
\widetilde{S}/\partial i\beta =0$ are obtained in the form (\ref{fi}) and (%
\ref{E}) as in the case of strong mixing, but the last equation and the DF
are different. The third equation in (\ref{delS}) is now%
\begin{equation}
\int d\mathbf{v}d\mathbf{r}\frac{\exp \left[ \frac{\eta (\tau
_{0})-\varepsilon (\mathbf{r},\mathbf{v})}{T}f_{0}(\tau _{0})\omega \right]
}{1+\int_{\Omega _{0}}d\tau _{0}\exp \left[ \frac{\eta (\tau )-\varepsilon (%
\mathbf{r},\mathbf{v})}{T}f_{0}(\tau _{0})\omega \right] }=1,  \label{N2}
\end{equation}%
and the equilibriun DF is%
\begin{equation}
\overline{f}_{LB}(\mathbf{r},\mathbf{v})=\frac{\int_{\Omega _{0}}d\tau
_{0}f_{0}(\tau _{0})\exp \left[ \frac{\eta (\tau _{0})-\varepsilon (\mathbf{r%
},\mathbf{v})}{T}f_{0}(\tau _{0})\omega \right] }{1+\int_{\Omega _{0}}d\tau
_{0}\exp \left[ \frac{\eta (\tau _{0})-\varepsilon (\mathbf{r},\mathbf{v})}{T%
}f_{0}(\tau _{0})\omega \right] }.  \label{LBDF}
\end{equation}%
The DF is presented in the form valid both for the natural units when $%
\omega =\mu (\Omega _{0})/k_{m}$ and for the dimensionless $\omega =1/k_{m}$
when $\mu (\Omega _{0})=1.$ The label LB indicates that DF (\ref{LBDF}) has
the form identical to the DF derived by Lynden-Bell. The relation of this
result with the assumed ergodicity without mixing is in line with
Lynden-Bell's idea of shape-preserving macroparticles. In Eqs. (\ref{N2})
and (\ref{LBDF}), the summation over $k$ is replaced by the integration over
the initial domain $\Omega _{0}$ occupied by the 6D liquid in the $\mu $
space. The quantity $\eta (\tau _{0})$ is now the function of $\tau _{0}\in
\Omega _{0}$\ and plays the role of a chemical potential for the particles
from the element $\sigma _{k}$ which contains the point $\tau _{0}$.

If $f_{0}(\tau _{0})=const,$ then the LBDF goes over into the Fermi-Dirac
distribution (\ref{DF1}). However, in both cases of unifom and nonuniform $%
f_{0}$ the physical meaning of the DF for ergodicity alone is very diferent
from DF for a mixing flow. In the last case, measuring density of liquid in
the $\mu $ space, one will obtain the time independent value $\overline{f}.$
In the ergodic case, the same measuremets will give results which are
different at different times, and the value $\overline{f}_{LB}(\mathbf{r},%
\mathbf{v})$ can be obtained only by computing the average over a very long
time. Thus, our calculation of the equilibrium DF in the pure ergodic case
is formal and must be undestood in the informal context above.

There is an interesting question, however, whether the spatial density,
which is obtained by averaging of $\overline{f}_{LB}(\mathbf{r},\mathbf{v})$
over $\mathbf{v}$, is time dependent or not. If on the integration over
velocities the time dependence is eliminated, then the results of measurment
of the spatial density can be predicted from the unpredictable $\mu $ space
density (\ref{LBDF}). This would make this DF of a high importance even in
the ergodic case. In a spatially uniform ergodic system, the average over
velocities is equivalent to the time average and therefore is time
independent. But this average is a constant independent of the spatial
coordinates. I am not aware of the analog of this theorem for a spatially
inhomogeneous system. Thus, a possible time independence of the spatial
density in the ergodic system without mixing seems to be interesting, but a
hypotetic issue.

\subsection{Maximum entropy state for incomplete mixing}

In the case of incomplete mixing with $M_{k}\sim N_{k}$ we have to use $%
Z_{i} $ in the form (\ref{ZM}). From the above consideration we know that,
in terms of the variabes $\varphi $, $\eta _{k}$, $T,$ the local partition
function $z_{i}$ (\ref{z}) has the form%
\begin{eqnarray}
z_{i} &=&\sum_{k}\exp \left( \frac{\eta _{k}-\varepsilon _{i}}{T}\frac{N_{k}%
}{M_{k}}\right)  \label{z1} \\
&=&\int_{\Omega _{0}}d\tau _{0}\exp \left[ \frac{\eta (\tau
_{0})-\varepsilon (\mathbf{r},\mathbf{v})}{T}\frac{f_{0}(\tau _{0})\omega }{%
M_{k}}\right] ,  \notag
\end{eqnarray}%
where $i\equiv \tau =(\mathbf{r},\mathbf{v})$. Now we use $Z_{i}$ (\ref{ZM})
with this $z_{i}$ in $\widetilde{S}$ (\ref{S1}) instead of $e^{\left\langle
\alpha N\right\rangle +c}$ and solve the statitical integral (\ref{Gamma2}).
The equations $\delta \widetilde{S}/\delta \psi =0$ and $\partial \widetilde{%
S}/\partial i\beta =0$ have the form (\ref{fi}) and (\ref{E}) as in the
cases of strong mixing and ergodicity. The equation $\partial \widetilde{S}%
/\partial (i\zeta _{k}N_{k})=0$ now is%
\begin{equation}
\int d\mathbf{v}d\mathbf{r}\frac{\int_{0}^{1}dx\exp \left[ \frac{\eta (\tau
_{0})-\varepsilon (\mathbf{r},\mathbf{v})}{T}f_{0}(\tau _{0})\omega x\right]
x^{-M_{k}x+1}}{1+\int_{0}^{1}dx\left\{ \int_{\Omega _{0}}d\tau _{0}\exp %
\left[ \frac{\eta (\tau _{0})-\varepsilon (\mathbf{r},\mathbf{v})}{T}\frac{%
f_{0}(\tau _{0})\omega }{M_{k}}\right] \right\} ^{M_{k}x}x^{-M_{k}x}}=1
\label{NN1}
\end{equation}%
and has to be satisfied for all $\tau _{0}\in \Omega _{0}$. The DF obtains
in the form
\begin{equation}
\overline{f}_{M}(\mathbf{r},\mathbf{v})=\frac{\int_{0}^{1}dx\left\{
\int_{\Omega _{0}}d\tau _{0}f_{0}(\tau _{0})\exp \left[ \frac{\eta (\tau
_{0})-\varepsilon (\mathbf{r},\mathbf{v})}{T}\frac{f_{0}(\tau _{0})\omega }{%
M_{k}}\right] \right\} ^{M_{k}x}x^{-M_{k}x+1}}{1+\int_{0}^{1}dx\left\{
\int_{\Omega _{0}}d\tau _{0}\exp \left[ \frac{\eta (\tau _{0})-\varepsilon (%
\mathbf{r},\mathbf{v})}{T}\frac{f_{0}(\tau _{0})\omega }{M_{k}}\right]
\right\} ^{M_{k}x}x^{-M_{k}x}},  \label{DFM}
\end{equation}%
\qquad

The DF $\overline{f}_{M}$ desribes the statistics of independent
subelements, which are enumerated by $\tau _{0}\in \sigma _{k}\subset \Omega
_{0}$ and which have the number of particles $N_{k}/M_{k}=f_{0}(\tau _{0})/M$
. We stress, however, that associating these subelements with macroparticles
would be incorrect as each subelement gets smeared over the accessible $\mu $
space.

Let us compare the equilibrium DF (\ref{DFM}) for the $M$ degree mixing with
the DF (\ref{DF1}) for strong mixing and the DF (\ref{LBDF}) for a pure
ergodic flow. As discussed in Sec. V.A, there are two relaxation processes
in the strong mixing flow. The fast relaxation prepares the perfect, energy
independent mixture of different density levels and the slow relaxation
distributes this mixture over the energy values. In the purely ergodic flow,
the mixture is not produced and the frequency of finding a given liquid
element at a point $\tau $\ depends solely on its energy at this point. In a
not perfectly mixing flow, the energy and mixing entropy compete. This is
well seen from the integrand in (\ref{DFM}) which, for $f_{0}\sim const$,
has the form
\begin{equation}
\exp \left( \frac{-\varepsilon }{T}N_{k}x-M_{k}x\ln x\right) ,
\label{compete}
\end{equation}%
clearly pointing at this competition.

To illustrate the peculiarity of the DF (\ref{DFM}), we resort to the
particular case of a constant liquid's density independent of $k,$ $%
N_{k}=f_{0}\omega =n_{0}=const$ (known as a "water bag" initial condition).
In this case $\overline{f}_{M}$ reduces to the following form:%
\begin{equation}
\overline{f}_{M}=f_{0}\frac{\int_{0}^{1}dx\exp \left[ \frac{\eta
-\varepsilon (\mathbf{r},\mathbf{v})}{T}n_{0}x\right] x^{-M_{k}x+1}}{%
1+\int_{0}^{1}dx\exp \left[ \frac{\eta -\varepsilon (\mathbf{r},\mathbf{v})}{%
T}n_{0}x\right] x^{-M_{k}x}}.  \label{DFM1}
\end{equation}%
As the DF $\overline{f}_{M}$ desribes the statistics of subelements with $%
n_{0}/M_{k}$ particles, it is instructive to compare this DF with the
Fermi-Dirac DF $f_{FD}$ for subelements which contain the same number of
particles. The two functions to be compared are%
\begin{eqnarray}
\overline{f}_{M} &=&\frac{\int_{0}^{1}dxe^{\phi x}x^{-M_{k}x+1}}{%
1+\int_{0}^{1}dxe^{\phi x}x^{-M_{k}x}},  \label{fM} \\
f_{FD} &=&\frac{e^{\phi /M_{k}}}{1+e^{\phi /M_{k}}},  \label{fFD}
\end{eqnarray}%
\ where $\phi =\frac{\eta -\varepsilon (\mathbf{r},\mathbf{v})}{T}n_{0}$. In
Fig. 1, the two DFs have been plotted vs $\phi $ for $M_{k}=500$. We see
that the dependence on the single argument $\phi $ in the DF $\overline{f}%
_{M}$ is much sharper than in the Fermi-Dirac DF for the same $M_{k}$. The
smaller $M_{k}$ the closer to one another the two functions become, but for
small $M$ the above theory looses its accuracy and is inapplicable for very
small $M_{k}\sim 1$. For very large $M_{k}$, the DF $\overline{f}_{M}$ goes
over into the DF (\ref{DF1}).

\begin{figure}[h]
\includegraphics[width=95mm]{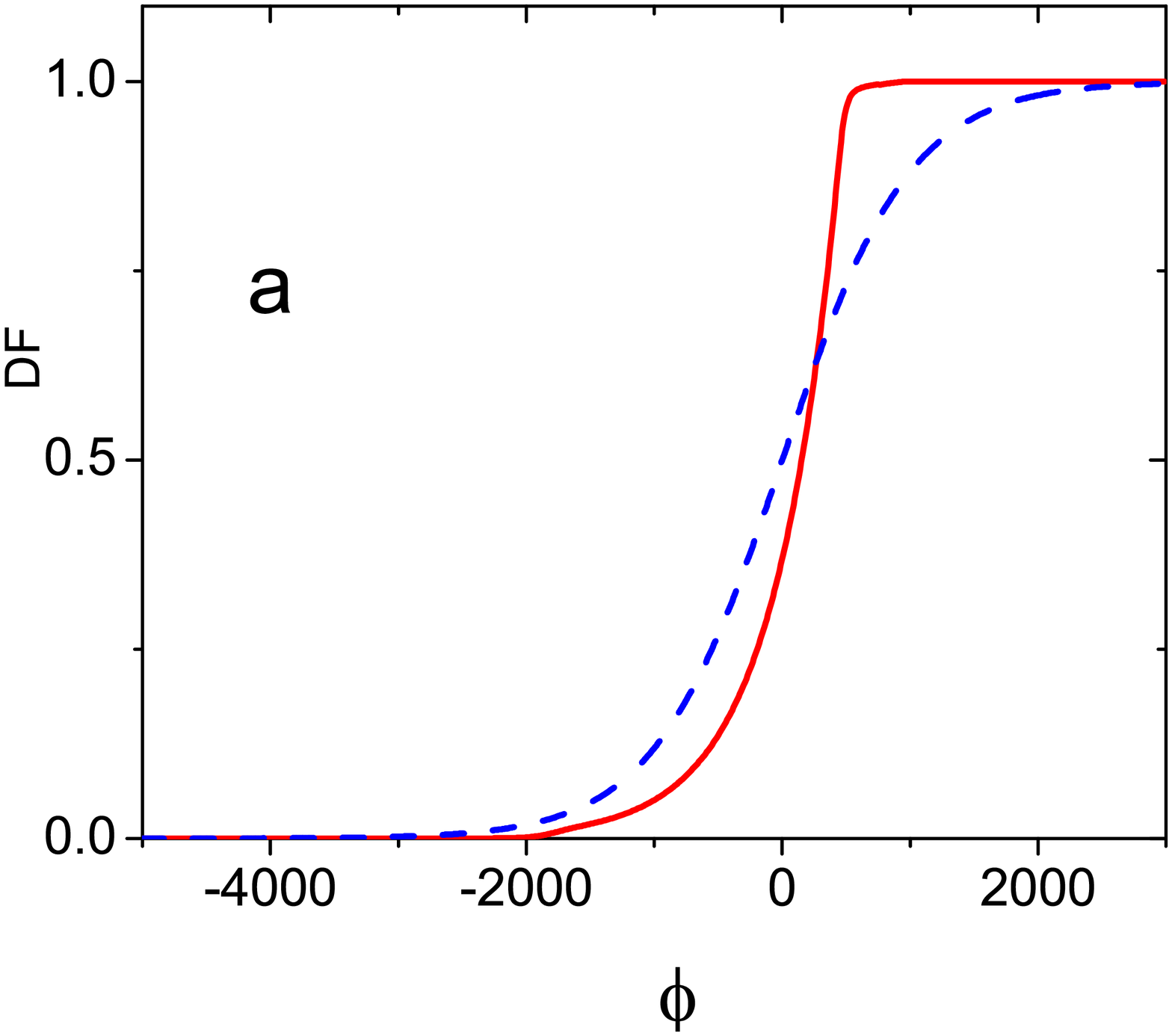} %
\includegraphics[width=95mm]{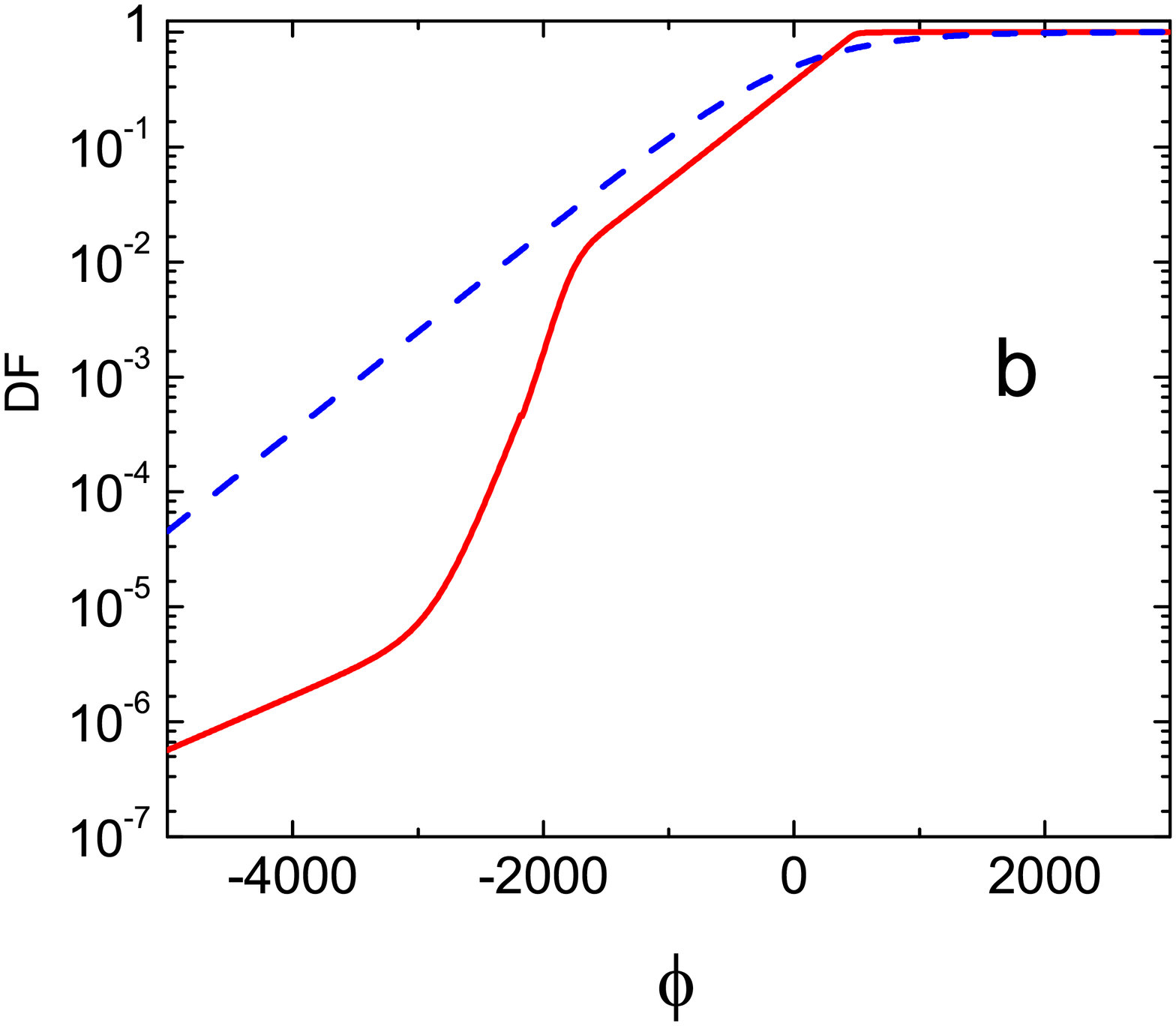}
\caption{Dependence of the incomplete mixing DF (\protect\ref{fM}) (solid
line) and the Fermi-Dirac DF (\protect\ref{fFD}) (dashed line) on $\protect%
\phi $ for mixing degree $M_{k}=500$. (a) - linear plot, (b) - semi
logarithmic plot.}
\label{fig:1}
\end{figure}
Further analysis of possible physical implications of the maximum entropy
states with incomplete mixing deserves a separate study.

\subsection{Some thermodynamics: Entropy, energy, temperature.}

The three statistics obtained are Fermi-Dirac-like which allows one to make
use of the well-known results from the theory of an ideal Fermi gas \cite%
{Landau}. Let us first realize that $\ln \Gamma $, the entropy $S$ of the
system, in the limit $N\rightarrow \infty $ is equal to the expression for
the exponent in formula (\ref{Gamma1}), calculated at the solution of the
equations (\ref{delS}). Indeed, on the $\psi $ integration the prefactor $%
\beta ^{-i_{m}/2}$ cancels out; for large $N$, the additional exponent
resulting from the Gauss integral over $d\beta D\zeta _{k}$, scales as $-\ln
N$ and is negligible as compared with the principal term which scales as $N$
(or even as $N^{2}$ as the potential energy of systems with a long-range
interaction is proportional to $N^{2}$). As a result, the entropy $S$ takes
the following form:%
\begin{equation}
S(E,N_{k})=\frac{E}{T}-\sum_{k=1}^{k_{m}}\frac{\eta _{k}N_{k}}{T}+\int d%
\mathbf{r}\left[ -\frac{(\mathbf{\nabla }\varphi )^{2}}{8\pi \gamma T}+\int d%
\mathbf{v}\ln (1+Z)\right] ,  \label{S2}
\end{equation}%
where $Z$ is defined in (\ref{ZM}).

The solution of the equations (\ref{delS}) depends on the parameter $E$ and
is a functional of the initial DF $f_{0}(\tau _{0})$ and the external field $%
\varphi _{ext}(\mathbf{r}).$ In the discrete version, the functional
dependence on $f_{0}$ is replaced by the dependence on the $k_{m}$
parameters $N_{k}:$
\begin{eqnarray}
T &=&T(E,N_{k},\varphi _{ext}),  \notag \\
\eta _{n} &=&\eta _{n}(E,N_{k},\varphi _{ext}),  \label{TT} \\
\varphi (\mathbf{r}) &=&\varphi (\mathbf{r}|E,N_{k},\varphi _{ext}),  \notag
\\
k,n &=&1,...,k_{m}.  \notag
\end{eqnarray}%
It is now easy to see that, by virtue of the equations (\ref{delS}), one has%
\begin{eqnarray}
\frac{\partial S}{\partial E} &=&\frac{1}{T},  \label{T} \\
\frac{\partial S}{\partial N_{k}} &=&-\frac{\eta _{k}}{T},  \label{eta} \\
\frac{\delta S}{\delta \varphi _{ext}(\mathbf{r})} &=&\frac{q\triangle
\varphi (\mathbf{r})}{4\pi \gamma },  \label{fi ext}
\end{eqnarray}%
where the expression in the rhs of (\ref{fi ext}) is just the spatial
density of particles $n(\mathbf{r})$ at point $\mathbf{r}$ times $q/\gamma $%
. Thus,
\begin{eqnarray}
TdS &=&dE-\sum_{k}\eta _{k}dN_{k}+\frac{q}{4\pi \gamma }\int d\mathbf{r}%
\delta \varphi _{ext}\triangle \varphi  \label{dS} \\
&=&dE-\int_{\Omega _{0}}d\tau _{0}\eta (\tau _{0})\delta f_{0}(\tau
_{0})+\left( q/\gamma \right) \int d\mathbf{r}n\delta \varphi _{ext},  \notag
\end{eqnarray}%
We see that $T$ can be associated with the global temperature and $\eta _{k}$
with the global chemical potential of the particles of the $k$-th kind
(which is determined by belonging to the liquid element with the density $%
N_{k}/\omega $). In an additive system with short-range interaction, these
two fundamental thermodynamic quantities are the characteristics of any
subsystem, but in our nonadditive system they can be attributed only to the
total system and in this sense are global. At the same time, the response to
a small variation of the external field, $\left( q/\gamma \right) n(\mathbf{r%
}),$ is local as in systems with a short-range interaction. The difference
is however that the density in systems with short-range interactions is
spatially dependent only in a nonzero external field whereas in systems with
long-range interaction $n(\mathbf{r})$ can be inhomogeneous even for zero
external field.

It is possible to simplify the expression for entropy. For convenience, let
us put $[\eta (\tau _{0})-q(\varphi +\varphi _{ext})]/T=u$. Then the general
expression for $Z$ (\ref{ZM}) is
\begin{equation}
Z=\int_{0}^{1}dx\left\{ \int_{\Omega _{0}}d\tau _{0}\exp \left[ \left( u-%
\frac{mv^{2}}{2T}\right) \frac{f_{0}(\tau _{0})\omega }{M_{k}}\right]
\right\} ^{M_{k}x}x^{-M_{k}x}.  \label{ZZZ}
\end{equation}%
Consider the integral
\begin{equation}
I=\int d\mathbf{v}\ln (1+Z).  \label{I1}
\end{equation}%
Integrating by parts, one obtains%
\begin{equation}
I=\frac{2}{3}\int d\mathbf{v}\frac{mv^{2}}{2T}\overline{f}_{M},  \label{I2}
\end{equation}%
where $\overline{f}_{M}$ is the DF (\ref{DFM}). Substituting this expression
in $S$, one gets%
\begin{equation}
TS=E-\sum_{k=1}^{k_{m}}\eta _{k}N_{k}+\int d\mathbf{r}\left[ -\frac{(\mathbf{%
\nabla }\varphi )^{2}}{8\pi \gamma }+\frac{2}{3}\int d\mathbf{v}\frac{mv^{2}%
}{2}\overline{f}_{M}\right] .  \label{S3}
\end{equation}%
It is now obvious that the spatial integral is equal to the total intrinsic
potential energy $U$ plus (2/3) of the total kinetic energy $K.$ Now we make
use of the 3D virial condition for our system \ \cite{Phys Rep 2014}. The
total energy $E=U+U_{ext}+K$ where $U_{ext}$ is the system's energy in the
external field:%
\begin{equation}
U_{ext}=\left( q/\gamma \right) \int d\mathbf{r}n\varphi _{ext}.
\label{Uext}
\end{equation}%
The most natural assumption is that the interaction of the particles with
the external field is also Coulomb-like and thus a homogeneous function of
order $-1$ \cite{Uext}. Then, in a stationary state, the 3D virial relation
tells one that $2K+U+U_{ext}=0$ so that $K=-E.$ Then $U+(2/3)K=(4/3)E-U_{ext}
$ and the entropy reduces to the following form:%
\begin{equation}
TS=\frac{7}{3}E-\int_{\Omega _{0}}d\tau _{0}\eta (\tau _{0})f_{0}(\tau
_{0})-U_{ext}.  \label{S4}
\end{equation}%
This formula is also valid for the ergodic case without mixing when the DF
is formally equal to the LBDF (\ref{LBDF}). In the case of strong mixing,
this relation reduces to%
\begin{equation}
TS=\frac{7}{3}E-\eta N-U_{ext}.  \label{S5}
\end{equation}%
To demonstrarte the effect of nonadditivity of our Coulomb-like long-range
interaction in 3 dimensions, we rewrite this formula in the following form:%
\begin{equation}
\eta N-\left( \frac{4}{3}E+2U_{ext}\right) =E-TS+U_{ext},  \label{Gibbs}
\end{equation}%
and compare it with the expression for the Gibbs free energy $\eta N=E-TS-PV$
(Gibbs-Duham equation) where the work of the external force $-PV$
corresponds to the energy in the external field $U_{ext}$ (work done by the
external source). We see that the correspondence between the two formulas is
violated by the term $-\left( \frac{4}{3}E+2U_{ext}\right) $ in the lhs of
our formula (\ref{Gibbs}). Very recently Latella et al \cite{Latella} showed
that the standard Gibbs-Duham equation is modified by the long-range
interaction also in systems which have already attained the final
equilibrium state due to the collisions. However, the direct comparison
between the collisional systems considered in \cite{Latella} and the
quasi-equilibrium collisionless systems considered in this paper cannot be
made. The reason is that, in collisional systems, the integrals over
velocities and coordinates are independent whereas, in collisionless
systems, they are not since the local consevation laws of Vlasov's equation
impose the connection between volumes of the liquid elements in the velocity
and coordinate spaces and, in the end, give rise to Eq.(\ref{S3}).

In the case of a Fermi gas, chemical potential at $T=0$ is the Fermi energy $%
\varepsilon _{F}$. As our statistics are Fermi-like, one obtains the
relation between the total energy and the chemical potential $\eta _{F}(\tau
_{0})$ of the fully degenerate 6D liquid:
\begin{eqnarray}
E &=&\frac{3}{7}\sum_{k=1}^{k_{m}}\eta _{F,k}N_{k}+U_{ext}  \label{EF1} \\
&=&\frac{3}{7}\int_{\Omega _{0}}d\tau _{0}\eta _{F}(\tau _{0})f_{0}(\tau
_{0})+U_{ext}.  \notag
\end{eqnarray}
In the case of strong mixing and $U_{ext}=0,$ this reduces to%
\begin{equation}
E=\frac{3}{7}\eta _{F}N,  \label{EF2}
\end{equation}%
while, for Fermi gas, this relation is $E=(3/5)\varepsilon _{F}N$. The
difference between our case of the long-range interaction and Fermi-gas is
however not just in the coefficients: $\varepsilon _{F\text{ }}$ is a
function of the particle density whereas $\eta _{F}$ depends on the total
number of particles $N$ (expectedly $\eta _{F}\sim N$). We note that in the
fully degenerate state the 6D liquid is stratified. The liquid is poured in
a 6D "glass" produced by the mean field potential and fills the 6D volume
which is exactly equal to the volume of the initial liquid's volume $\mu
(\Omega _{0}).$ At the very bottom is the liquid element with the maximum $%
N_{k}$ which occupies the volume $\omega $; the layer of liquid with the
next large $N_{k}$ lies over the liquid with the largest $N_{k},$ and so on.
The quantity $\eta _{F,k}$ is the energy of a particle at the top of the
layer with $N_{k}.$ It is worth noting that this state is the absolute
minimum energy state and can be realized without any mixing just via
precipitation of havier fractions closer to the bottom of the potential well.

\section{Conclusion.}

Almost fifty years ago Linden-Bell assumed that the 6D liquid described by
the Vlasov equation can relax to some maximum entropy state \cite{LB}. This
required a very strong and highly idealized assumption of a "good" mixing in
the liquid's phase space. Moreover, any mixing implies that liquid elements
spread over the accessible $\mu $ space and thus do not preserve their
initial shapes. In spite of that Linden-Bell modeled the liquid by a large
set of independent shape-preserving macroparticles in an obvious
contradiction to the mixing character of the flow. Nevertheless, this result
has had a profound influence both on the astrophysics and statistical
mechanics. Over the years, there has been a number of papers where the
equilibrium maximum entropy state was addressed by means of different
technics and in relations to different systems with long-range interaction
\cite{Phys Lett,Tech 1,Tech 2,Tech 3,Tech 3a,Tech 4,Chav book}. But the main
peculiarity of a liquid motion, a flow, which must be mixing if we expect
relaxation to some statistical eqilibrium and thus making very small liquid
elements spread over the entire space, has not been incorporated. In other
words, the actual 6D liquid has not been adequately described. In this
paper, based on the ideas of the ergodic theory, I developed an equilibrium
statistical approach to an actual 6D liquid in the $\mu $ space and
incorporated all the integrals of motion of the Vlasov equation, associated
with the point-wise incompressibility of the liquid. I specified the
assumption of mixing. It can be very strong, practically a K mixing when all
points of a liquid are statistically independent. Or mixing can be
incomplete, e.g., approximately one independent liquid element per particle.
And mixing can be very weak, equivalent to pure ergodicity when the number
of independent liquid elements is very small. In the case of mixing, the
flow relaxes to certain equilibrium state which depends on the mixing
strength. For a very strong mixing, this state is given exactly by the
Fermi-Dirac DF (\ref{DF1}), in which different levels of the initial DF are
perfectly mixed and do not appear individually. For incomplete mixing,
different levels manifest themselves and, in addition, the DF (\ref{DFM})
depends on the density of the statistically independent liquid elements. It
is found that the LBDF (\ref{LBDF}) formally corresponds to a pure ergodic
flow which does not result in relaxation to a time independent state. The
phase space of Vlasov's liquid where its macrostate is given by a single
point (in contrast to the $\mu $ space where the liquid is desribed by its
multipoint shape) is introduced. The integrals of motions are enumerated by
the points of the support of the initial DF. The crucial result of the paper
is expressed by the formulae (\ref{Wa}) and (\ref{hk}) - (\ref{h}) which
give the number of microscopic states in a given macroscopic state of a
liquid. This result is based on the two ideas: a) introduction of the
ensemble of phase subspaces which is a continuous analog of accounting for
distributions of particles realized on subspaces of dimensions smaller than
that of the total space of states; b) connecting microscopic states in a
given macrostate with the sequences produced by appropriate Bernulli mapping
for which the number of observable sequences (i.e., those with nonvanishing
probability) is given by the Shannon-MaMillan-Breiman theorem. The relation
between the entropy, total energy, and the initial DF is derived. The result
of this paper clarifies status of maximum entropy states which, under any
possible assumptions of the chaoticity of Vlasov's flow, can, in principle,
be achieved. The introduced idea of incomplete relaxation connected to the
number of statistically independent liquid elements is probably just one
approach and does not exhaust all possibilities. Further studies in this
direction can clarify the status of the incomplete relaxation and the DF (%
\ref{DFM}) in a more general context. The presented approach can be viewed
as a step in development of the statistical mechanical technics adequate for
dealing with continuous, distributed, and liquid-like systems. Specific
effects that may be derived from the DF (\ref{DFM}) will be studied in the
future.

\emph{Acknowledgement.}

I am grateful to my late friend and teacher, fine experimentalist Vitali P.
Kovalenko, who, once upon a time, referred me to Lynden-Bell's paper.


\begin{thebibliography}{99}
\bibitem{Phys Rep 1990} T. Padmanabhan, Phys. Rep. \textbf{188}, 285 (1990).

\bibitem{book} \textit{Dynamics and Thermodynamics of Systems with
Long-Range Interaction}, Edited: T. Dauxois, S. Ruffo, E. Arimondo, M.
Wilkens, Lecture Notes in Physics, Vol. 602 (Springer, Berlin, 2002).

\bibitem{Phys Rep 2009} A. Campa, T. Dauxois, and S. Ruffo, Phys. Rep.
\textbf{480}, 57 (2009)

\bibitem{Phys Rep 2014} Y. Levin, R. Pakter, F. B. Rizzato, T. N. Teles, F.
P. C. Benetti, Phys. Rep. \textbf{535}, 1 (2014).

\bibitem{1.7 1} M. Antoni and S. Ruffo, Phys. Rev. E \textbf{52}, 2361
(1995).

\bibitem{1.7 2} V. Latora, A. Rapisarda, S. Ruffo, Phys. Rev. Lett. \textbf{%
80}, 792 (1998).

\bibitem{1.7 3} V. Latora, A. Rapisarda, C. Tsallis, Phys. Rev. E \textbf{64}%
, 056134 (2001).

\bibitem{1.7 4} Y.Y. Yamaguchi, J. Barr, F. Boucheta, T. Dauxoisc, S. Ruffo,
Physica (Amsterdam) \textbf{337A}, 36 (2004).

\bibitem{1.7 5} K. Jain, F. Bouchet, D. Mukamel, J. Stat. Mech.: Theory and
Experiment, P11008 (2007).

\bibitem{1.7 6} Y. Levin, R. Pakter, F. Rizzato, Phys. Rev. E \textbf{78}
021130 (2008).

\bibitem{LB} D. Lynden-Bell, Mon. Not. Roy. Astr. Soc. \textbf{136},101
(1967).

\bibitem{Fignja PRL} T. M. R. Filho, A. Figueiredo, and M. A. Amato, Phys.
Rev. Lett. \textbf{95}, 190601 (2005).

\bibitem{Phys Lett} V.M. Pergamenshchik, Phys. Lett. \textbf{113}A, 225
(1985).

\bibitem{Tech 1} J. Miller, Phys. Rev. Lett. \textbf{65}, 2137 (1990).

\bibitem{Tech 2} R. Robert and J. Sommeria, J. Fluid Mech. \textbf{229}, 291
(1991).

\bibitem{Tech 3} J. Miller, P.B. Weichman, and M.C. Cross, Phys. Rev. A
\textbf{45}, 2328 (1992).

\bibitem{Tech 3a} P.H. Chavanis, Statistical mechanics of violent relaxation
in stellar systems, Proceedings of the Conference on Multiscale Problems in
Science and Technology, (Springer, 2002).

\bibitem{Chav book} P.-H. Chavanis, Statistical mechanics of two-dimensional
vortices and stellar systems, in \cite{book}.

\bibitem{Tech 4} P.-H. Chavanis, Physica A \textbf{359,} 177 (2006).

\bibitem{LB critics} I. Arad and D. Lynden-Bell, Mon. Not. R. Astron. Soc.
\textbf{361}, 385 (2005).

\bibitem{Prais 1} A. Antoniazzi, D. Fanelli, J. Barr\'{e}, P.-H. Chavanis,
T. Dauxois, S. Ruffo, Phys. Rev. E \textbf{75}, 011112 (2007).

\bibitem{Prais 2} Y. Levin, R. Pakter, and F. B. Rizzato, Phys. Rev. E
\textbf{78} 021130 (2008).

\bibitem{Prais 3} P. de Buyl, D. Mukamel, and S. Ruffo, Phil. Trans. R. Soc.
A \textbf{369}, 439 (2011).

\bibitem{Critics 2} R. Pakter and Y. Levin, Phys. Rev. Lett. \textbf{110},
14061 (2013).

\bibitem{Critics 3} T.M.R. Filho, M.A. Amato, and A. Figueiredo, Phys. Rev.
E \textbf{85,} 62103 (2012).

\bibitem{Nakamura} T.K. Nakamura, Astrophys. J. \textbf{531} (2000) 739.

\bibitem{PRE2009} V.M. Pergamenshchik, Phys. Rev. E \textbf{79,} 011407
(2009); \textbf{85,} 021403 (2012).

\bibitem{Balescu} R. Balescu, \textit{Equilibrium and Nonequilibrium
Statistical Mechanics} (John Wiley, NY, 1975).

\bibitem{Reichl} L. Reichl, \textit{A Modern Course of Statistical Physics}
(Wiley, 1998).

\bibitem{Bil} P. Billingsley,\textit{\ Ergodic Theory and Information} (John
Wiley, NY, 1965).

\bibitem{Sinai} I.P. Kornfeld, S.V. Fomin, and Y.G. Sinai, \textit{Ergodic
theory }(Springer, 1982).

\bibitem{Ideal gas} For instance, the distribution of an ideal gas over its
accessible $6N$ dimensional phase space is homogeneous, but the
Maxwell-Boltzmann DF, which is the DF in its 6D $\mu $ space, is
inhomogeneous.

\bibitem{Complex Gauss} Usually, the Habbard-Stratonovich transformation is
based on a real Gauss integral. However, our formula (\ref{HS}) is based on
a complex Gauss integral, see, e.g., \cite{Bogoljubov}, to incorporate both
positive and negative sign of $\gamma $ in the interaction potential (\ref{w}%
).

\bibitem{Bogoljubov} N.N. Bogoliubov and D.V. Shirkov, \textit{Introduction
to the Theory of Quantized Fields} (Interscience, New York, 1959).

\bibitem{Landau} L.D. Landau and E.M. Lifshitz, \textit{Statistical Physics,
Part 1} (Pergamon Press, Oxford, 1980).

\bibitem{Uext} Of course, it is difficult to imagine how $E$ can be fixed in
an external field which can change the system's energy. The external field
for such "isolated system" plays the role analogous to external pressure. In
this sense, the inclusion of an external field is somewhat formal.

\bibitem{Latella} I. Latella, A. P\'{e}rez-Madrid, A. Campa, L. Casetti, and
S. Ruffo, Phys. Rev. Lett. \textbf{114}, 230601 (2015).
\end{thebibliography}
\end{document}